\documentclass[
  journal=pasa,
  manuscript=research-paper, 
  year=2025,
]{cup-journal}

\makeatletter
\def\@maketitle{%
  \setlength\parindent{\z@}
  \ifnum\cup@author@cnt<\z@\relax
    \cup@warning{No authors defined: At least one author is required}%
  \fi
  \newpage\vspace*{\dimexpr-\headsep-\headheight\relax}
  \parbox[b]{\dimexpr\textwidth-26mm\relax}{%
    \firstheadsize{\journalnamefont\cup@journal@name} 
    (\cup@year), {\volumefont\cup@vol}, \thepage--\pageref{LastPage}
    \par doi:{\cup@doi}}\hfill\includegraphics[width=26mm]{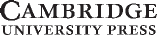}\par
  \vspace*{\baselineskip}
  {\sffamily\bfseries\color{structure@color}\MakeUppercase{\convertchar{\cup@manuscript}{-}{ }}\par}
  \vspace*{\cup@space@pre@title}%
        {%
          \titlefont
          \titlesize
          \let\@fnsymbol\cup@author@fnsymbol
          \let\footnote@org\footnote
          \let\footnote\cup@title@footnote
          \cup@maketitle@suppinfo \@title
          \cup@title@footnote@check
          \global\cup@footnote@cnt\c@footnote
          \@maketitle@title@hook
          \par
        }%
        \vspace*{\cup@space@post@title}%
        {%
          \authorsize
          \authorfont
          \frenchspacing
          \cup@author@list 
          \par
        }%
        \vspace*{\cup@space@post@author}%
        {%
          \emailsize
          \emailfont
          \ifcup@email
            \expandafter\cup@contact@details
            \par
            \vspace*{\cup@space@post@email}%
          \fi
        }%
        {%
          \datesize
          \datefont
          \ifcup@dates
            (Received \cup@recvd; revised \cup@revd; accepted \cup@accptd; first published online \cup@publd)
            \vspace*{\cup@space@post@date}%
          \fi
          \ifcup@copyright
            \bgroup\let\thefootnote\relax
            \footnote{\cup@copyright@notice}
            \egroup
          \fi
        }%
}
\makeatother

\newcommand{\printaffiliations}{%
  \section*{Affiliations}
  \affilsize
  \affilfont
  \begingroup
  \newcount\affilcount
  \affilcount=0
  \loop\advance\affilcount by 1
    \typeout{Checking affiliation number: \the\affilcount} 
    \expandafter\ifx\csname @address@\romannumeral\affilcount\endcsname\relax
      \typeout{Affiliation \the\affilcount not found, exiting loop.}
      \break
    \else
      \noindent\textbf{\textsuperscript{\the\affilcount}} 
      \csname @address@\romannumeral\affilcount\endcsname\par 
    \fi
  \ifnum\affilcount<\cup
  \repeat
  \endgroup
}

\usepackage{amsmath}
\usepackage{dcolumn}
\usepackage{rotating}
\usepackage{booktabs}

\usepackage{orcidlink}

\usepackage{hyperref}

\hypersetup{colorlinks,citecolor=blue,linkcolor=blue,urlcolor=blue}

\def\lesssim{\ifmmode\stackrel{<}{_{\sim}}\else$\stackrel{<}{_{\sim}}$\fi} 
\def\gtrsim{\ifmmode\stackrel{>}{_{\sim}}\else$\stackrel{>}{_{\sim}}$\fi} 
\def\HI{H\,{\sc i}}
\def\HII{H\,{\sc ii}}
\def\degr{^\circ}
\def\fdg{\hbox{$.\mkern-4mu^\circ$}}

\def\farcs{\hbox{$''\!\!.\,$}}

\usepackage{aas-macros}

\usepackage{microtype,siunitx,booktabs}
\title{The Polarisation Sky Survey of the Universe's Magnetism (POSSUM): \linebreak Science Goals and Survey Description}

\author{B. M. Gaensler \orcidlink{0000-0002-3382-9558}}
\affiliation{Department of Astronomy and Astrophysics, University of California Santa Cruz, 1156 High Street, Santa Cruz, CA 95064, USA}
\alsoaffiliation{Dunlap Institute for Astronomy and Astrophysics, University of Toronto, 50 St. George Street, Toronto, ON M5S 3H4, Canada}
\alsoaffiliation{David A. Dunlap Department of Astronomy and Astrophysics, University of Toronto, 50 St. George Street, Toronto, ON M5S 3H4, Canada}

\author{G. H. Heald \orcidlink{0000-0002-2155-6054}}
\affiliation{Australia Telescope National Facility, CSIRO, Space and Astronomy, PO Box 1130, Bentley WA 6102, Australia}
\alsoaffiliation{SKA Observatory, SKA-Low Science Operations Centre, 26 Dick Perry Avenue, Kensington WA 6151, Australia}
\email[G. H. Heald]{george.heald@skao.int}

\author{N. M. McClure-Griffiths \orcidlink{0000-0003-2730-957X}}
\affiliation{Research School of Astronomy \& Astrophysics, The Australian National University, Canberra ACT 2611, Australia}

\author{C. S. Anderson
\orcidlink{0000-0002-6243-7879}}
\affiliation{Research School of Astronomy \& Astrophysics, The Australian National University, Canberra ACT 2611, Australia}

\author{C. L. Van Eck\orcidlink{0000-0002-7641-9946}}
\affiliation{Research School of Astronomy \& Astrophysics, The Australian National University, Canberra ACT 2611, Australia}
\alsoaffiliation{Dunlap Institute for Astronomy and Astrophysics, University of Toronto, 50 St. George Street, Toronto, ON M5S 3H4, Canada}

\author{J. L. West \orcidlink{0000-0001-7722-8458}}
\affiliation{National Research Council Canada, Herzberg Research Centre for Astronomy and Astrophysics, Dominion Radio Astrophysical Observatory, PO Box 248, Penticton, BC V2A 6J9, Canada}
\alsoaffiliation{Dunlap Institute for Astronomy and Astrophysics, University of Toronto, 50 St. George Street, Toronto, ON M5S 3H4, Canada}

\author{A. J. M. Thomson \orcidlink{0000-0001-9472-041X}}
\affiliation{Australia Telescope National Facility, CSIRO, Space and Astronomy, PO Box 1130, Bentley WA 6102, Australia}

\author{J. P. Leahy \orcidlink{0000-0003-2514-9592}}
\affiliation{Jodrell Bank Centre for Astrophysics, Department of Physics and Astronomy, University of Manchester, Manchester M13 9PL, UK}

\author{L. Rudnick}
\affiliation{Minnesota Institute for Astrophysics, University of Minnesota, 116 Church Street SE, Minneapolis, MN 55455, USA}

\author{Y. K. Ma \orcidlink{0000-0003-0742-2006}}
\affiliation{Research School of Astronomy \& Astrophysics, The Australian National University, Canberra ACT 2611, Australia}

\author{Takuya Akahori \orcidlink{0000-0001-9399-5331}}
\affiliation{Mizusawa VLBI Observatory, National Astronomical Observatory of Japan, 2-21-1, Osawa, Mitaka, Tokyo 181-8588, Japan
}

\author{G. G\"urkan \orcidlink{0000-0002-9777-1762}}
\affiliation{Centre for Astrophysics Research, University of Hertfordshire, College Lane, Hatfield AL10 9AB, UK}
\alsoaffiliation{Australia Telescope National Facility, CSIRO, Space and Astronomy, PO Box 1130, Bentley WA 6102, Australia}

\author{T. L. Landecker \orcidlink{0000-0003-1455-2546}}
\affiliation{National Research Council Canada, Herzberg Research Centre for Astronomy and Astrophysics, Dominion Radio Astrophysical Observatory, PO Box 248, Penticton, BC V2A 6J9, Canada}

\author{S. A. Mao}
\affiliation{Max Planck Institute for Radio Astronomy, Auf dem H\"ugel 69, 53121 Bonn, Germany}

\author{S. P. O'Sullivan \orcidlink{0000-0002-3968-3051}}
\affiliation{Departamento de F\'{i}sica de la Tierra y Astrof\'{i}sica \& IPARCOS-UCM, Universidad Complutense de Madrid, 28040 Madrid, Spain}

\author{W. Raja}
\affiliation{Australia Telescope National Facility, CSIRO Space and Astronomy, PO Box 76, Epping, NSW 1710, Australia}

\author{X. Sun \orcidlink{0000-0002-3464-5128}}
\affiliation{School of Physics and Astronomy, Yunnan University, Kunming 650500, China}

\author{T. Vernstrom}
\affiliation{Australia Telescope National Facility, CSIRO, Space and Astronomy, PO Box 1130, Bentley WA 6102, Australia}
\alsoaffiliation{International Centre for Radio Astronomy Research, University of Western Australia, 7 Fairway, Crawley, WA 6009, Australia}

\author{Lerato Baidoo \orcidlink{0000-0003-0520-0696}}
\affiliation{Dunlap Institute for Astronomy and Astrophysics, University of Toronto, 50 St. George Street, Toronto, ON M5S 3H4, Canada}

\author{Ettore Carretti \orcidlink{0000-0002-3973-8403}}
\affiliation{INAF -- Istituto di Radioastronomia, via P. Gobetti 101, 40129 Bologna, Italy}

\author{A. R. Taylor \orcidlink{0000-0001-9885-0676}}
\affiliation{Inter-University Institute for Data Intensive Astronomy, University of Cape Town, Rondebosch, 7701, South Africa}
\alsoaffiliation{Department of Astronomy, University of Cape Town, Rondebosch, Cape Town, 7701, South Africa}
\alsoaffiliation{Department of Physics and Astronomy, University of the Western Cape, Bellville, Cape Town, 7535 South Africa}

\author{A.G. Willis}
\affiliation{National Research Council Canada, Herzberg Research Centre for Astronomy and Astrophysics, Dominion Radio Astrophysical Observatory, PO Box 248, Penticton, BC V2A 6J9, Canada}

\author{Erik Osinga \orcidlink{0000-0002-5815-8965}}
\affiliation{Dunlap Institute for Astronomy and Astrophysics, University of Toronto, 50 St. George Street, Toronto, ON M5S 3H4, Canada}

\author{J. D. Livingston}
\affiliation{Max Planck Institute for Radio Astronomy, Auf dem H\"ugel 69, 53121 Bonn, Germany}

\author{E. L. Alexander \orcidlink{0000-0002-6249-614X}}
\affiliation{Jodrell Bank Centre for Astrophysics, Department of Physics and Astronomy, University of Manchester, Manchester M13 9PL, UK}
\alsoaffiliation{School of Physics \& Astronomy, University of Leeds, Leeds, LS2 9JT, UK}

\author{David Alonso-L\'opez \orcidlink{0000-0001-9006-0725}}
\affiliation{Departamento de F\'{i}sica de la Tierra y Astrof\'{i}sica \& IPARCOS-UCM, Universidad Complutense de Madrid, 28040 Madrid, Spain}

\author{A. D. Amaral}
\affiliation{David A. Dunlap Department of Astronomy and Astrophysics, University of Toronto, 50 St. George Street, Toronto, ON M5S 3H4, Canada}
\alsoaffiliation{Dunlap Institute for Astronomy and Astrophysics, University of Toronto, 50 St. George Street, Toronto, ON M5S 3H4, Canada}

\author{T. An \orcidlink{0000-0003-4341-0029}}
\affiliation{Shanghai Astronomical Observatory, CAS, 80 Nandan Road, Shanghai 200030, China}
\alsoaffiliation{100101 Key Laboratory of Radio Astronomy and Technology (Chinese Academy of Sciences), A20 Datun Road, Chaoyang District, Beijing, 100101, P. R. China}

\author{Andrea Bracco \orcidlink{0000-0003-0932-3140}}
\affiliation{INAF – Osservatorio Astrofisico di Arcetri, Largo E. Fermi 5, 50125 Firenze, Italy}
\alsoaffiliation{Laboratoire de Physique de l'Ecole Normale Sup\'erieure, ENS, Universit\'e PSL, CNRS, Sorbonne Universit\'e, Universit\'e de Paris, F-75005 Paris, France}

\author{S. Bradbury \orcidlink{0000-0001-9261-885X}}
\affiliation{Research School of Astronomy \& Astrophysics, The Australian National University, Canberra ACT 2611, Australia}

\author{Marcus Br\"uggen \orcidlink{0000-0002-3369-7735}}
\affiliation{Hamburger Sternwarte, University of Hamburg, Gojenbergsweg 112, 21029 Hamburg, Germany}

\author{Chakali Eswaraiah}
\affiliation{Department of Physics, Indian Institute of Science Education and Research (IISER), Tirupati 517507, India}

\author{Torsten En\ss lin}
\affiliation{Max Planck Institute for Astrophysics, Karl-Schwarzschild-Straße 1, 85748 Garching, Germany}

\author{T. J. Galvin}
\affiliation{Australia Telescope National Facility, CSIRO, Space and Astronomy, PO Box 1130, Bentley WA 6102, Australia}

\author{Marijke Haverkorn \orcidlink{0000-0002-5288-312X}}
\affiliation{Department of Astrophysics / IMAPP, Radboud University Nijmegen, PO Box 9010, 6500 GL Nijmegen, The Netherlands}

\author{A. M. Hopkins}
\affiliation{School of Mathematical and Physical Sciences, 12 Wally's Walk, Macquarie University, NSW 2109, Australia}

\author{Sebastian Hutschenreuter \orcidlink{0000-0002-6952-9688}}
\affiliation{University of Vienna, Department of Astrophysics, Türkenschanzstrasse
17, 1180 Vienna, Austria}

\author{Shinsuke Ideguchi}
\affiliation{National Astronomical Observatory of Japan, 2-21-1 Osawa, Mitaka, Tokyo 181-8588, Japan}

\author{S. Jaswanth}
\affiliation{Max Planck Institute for Radio Astronomy, Auf dem H\"ugel 69, 53121 Bonn, Germany}

\author{S. Lyla Jung \orcidlink{0000-0001-5512-3735}}
\affiliation{Astrophysics, Denys Wilkinson Building, Department of Physics, University of Oxford, Keble Road, Oxford OX1 3RH, United Kingdom}

\author{J. F. Kaczmarek \orcidlink{0000-0003-4810-7803}}
\affiliation{Australia Telescope National Facility, CSIRO Space \&\ Astronomy, Parkes Observatory, P.O. Box 276, Parkes, NSW 2870, Australia}

\author{Roland Kothes \orcidlink{0000-0001-5953-0100}}
\affiliation{National Research Council Canada, Herzberg Research Centre for Astronomy and Astrophysics, Dominion Radio Astrophysical Observatory, PO Box 248, Penticton, BC V2A 6J9, Canada}

\author{Sanja Lazarevi\'{c} \orcidlink{0000-0001-6109-8548}}
\affiliation{Western Sydney University, Locked Bag 1797, Penrith South DC, NSW 2751, Australia}
\alsoaffiliation{Australia Telescope National Facility, CSIRO Space and Astronomy, PO Box 76, Epping, NSW 1710, Australia}
\alsoaffiliation{Astronomical Observatory, Volgina 7, 11060 Belgrade, Serbia}

\author{Denis Leahy \orcidlink{0000-0002-4814-958X}}
\affiliation{Department of Physics and Astronomy, The University of Calgary, 2500 University Drive NW, Calgary AB T2N 1N4, Canada}

\author{Francesca Loi \orcidlink{0000-0002-8627-6627}}
\affiliation{INAF - Osservatorio Astronomico di Cagliari, via della scienza 5, Selargius, Italy}

\author{Joshua R. Marvil}
\affiliation{National Radio Astronomy Observatory, P.O. Box O, Socorro, NM 87801, USA}

\author{Ray Norris \orcidlink{0000-0002-4597-1906}}
\affiliation{Australia Telescope National Facility, CSIRO Space and Astronomy, PO Box 76, Epping, NSW 1710, Australia}
\alsoaffiliation{Western Sydney University, Locked Bag 1797, Penrith South DC, NSW 2751, Australia}

\author{Ayush Pandhi}
\affiliation{David A. Dunlap Department of Astronomy and Astrophysics, University of Toronto, 50 St. George Street, Toronto, ON M5S 3H4, Canada}

\author{Jason M. Price \orcidlink{0000-0001-7013-0562}}
\affiliation{Research School of Astronomy \& Astrophysics, The Australian National University, Canberra ACT 2611, Australia}

\author{C. J. Riseley \orcidlink{0000-0002-3369-1085}}
\affiliation{Astronomisches Institut der Ruhr-Universit\"{a}t Bochum (AIRUB), Universit\"{a}tsstra{\ss}e 150, 44801 Bochum, Germany}

\author{P. Ryder}
\affiliation{Jodrell Bank Centre for Astrophysics, Department of Physics and Astronomy, University of Manchester, Manchester M13 9PL, UK}

\author{Amit Seta \orcidlink{0000-0001-9708-0286}}
\affiliation{Research School of Astronomy \& Astrophysics, The Australian National University, Canberra ACT 2611, Australia}

\author{Vasundhara Shaw}
\affiliation{Jodrell Bank Centre for Astrophysics, Department of Physics and Astronomy, University of Manchester, Manchester M13 9PL, UK}

\author{A. X. Shen}
\affiliation{Australia Telescope National Facility, CSIRO, Space and Astronomy, PO Box 1130, Bentley WA 6102, Australia}

\author{C. Sobey \orcidlink{0000-0002-8950-7873}}
\affiliation{Australia Telescope National Facility, CSIRO, Space and Astronomy, PO Box 1130, Bentley WA 6102, Australia}
\alsoaffiliation{SKA Observatory, SKA-Low Science Operations Centre, 26 Dick Perry Avenue, Kensington WA 6151, Australia}

\author{J. Stil \orcidlink{0000-0003-2623-2064}}
\affiliation{Department of Physics and Astronomy, The University of Calgary, 2500 University Drive NW, Calgary AB T2N 1N4, Canada}

\author{Chiara Stuardi \orcidlink{0000-0003-1619-3479}}
\affiliation{INAF -- Istituto di Radioastronomia, via P. Gobetti 101, 40129 Bologna, Italy}

\author{Gupta Upasana}
\affiliation{Indian Institute of Science Education and Research (IISER), Tirupati 517619, India}

\author{Shannon Vanderwoude \orcidlink{0009-0004-7773-1618}}
\affiliation{Dunlap Institute for Astronomy and Astrophysics, University of Toronto, 50 St. George Street, Toronto, ON M5S 3H4, Canada}

\author{Velibor Velovi\'{c} \orcidlink{0000-0002-0416-3267}}
\affiliation{Western Sydney University, Locked Bag 1797, Penrith South DC, NSW 2751, Australia}

\doi{10.1017/pas.\the\year.xxx}

\received {dd Mmm YYYY}
\revised  {dd Mmm YYYY}
\accepted {dd Mmm YYYY}
\published{dd Mmm YYYY}


\begin{document}

\begin{abstract}

The Australian SKA Pathfinder (ASKAP) offers powerful new capabilities for studying the polarised and magnetised Universe at radio wavelengths. In this paper, we introduce the Polarisation Sky Survey of the Universe's Magnetism (POSSUM), a groundbreaking survey with three primary objectives: (1) to create a comprehensive Faraday rotation measure (RM) grid of up to one million compact extragalactic sources across the southern $\sim50$ per cent of the sky (20,630\,deg$^2$); (2) to map the intrinsic polarisation and RM properties of a wide range of discrete extragalactic and Galactic objects over the same area; and (3) to contribute interferometric data with excellent surface brightness sensitivity, which can be combined with single-dish data to study the diffuse Galactic interstellar medium. Observations for the full POSSUM survey commenced in May 2023 and are expected to conclude by mid-2028. POSSUM will achieve an RM grid density of around 30--50 RMs per square degree with a median measurement uncertainty of $\sim$1~rad~m$^{-2}$. The survey operates primarily over a frequency range of 800--1088 MHz, with an angular resolution of $20''$ and a typical RMS sensitivity in Stokes $Q$ or $U$ of 18~$\mu$Jy~beam$^{-1}$. Additionally, the survey will be supplemented by similar observations covering 1296--1440 MHz over 38 per cent of the sky. POSSUM will enable the discovery and detailed investigation of magnetised phenomena in a wide range of cosmic environments, including the intergalactic medium and cosmic web, galaxy clusters and groups, active galactic nuclei and radio galaxies, the Magellanic System and other nearby galaxies, galaxy halos and the circumgalactic medium, and the magnetic structure of the Milky Way across a very wide range of scales, as well as the interplay between these components. This paper reviews the current science case developed by the POSSUM Collaboration and provides an overview of POSSUM's observations, data processing, outputs, and its complementarity with other radio and multi-wavelength surveys, including future work with the SKA.

\end{abstract}

\section{INTRODUCTION}
\label{sec:intro}

Magnetic fields pervade the cosmos, influencing a wide variety of phenomena across a vast range of scales---from planetary habitability and interstellar turbulence to the cooling of galaxy clusters and the propagation of cosmic rays. Magnetism, alongside gravity, drives the dynamics of baryonic matter and is a key component of the feedback mechanisms that regulate gas inflow and outflow in galaxies. Despite their ubiquity, however, our observational understanding of cosmic magnetic fields remains surprisingly limited. As reviewed in \S\ref{sec_sci}, we have yet to establish the magnetic geometries of many classes of celestial objects. We do not know how magnetism first originated in the Universe, nor do we understand in depth how the first magnetic fields were organized and amplified to produce the fields observed today \citep{2016RPPh...79g6901S}.

Astrophysical magnetic fields can be studied in various ways \citep[see Chapter 2 of][]{2016cmf..book.....K}, but the most versatile methods involve observations of linearly polarised radio synchrotron emission. The widespread presence of bright polarized sources across the Universe, along with the ability of radio waves to penetrate environments that are often inaccessible at other wavelengths due to extinction or absorption, makes radio polarimetry a powerful tool for exploring astrophysical and cosmic magnetism over vast distances and within complex environments. 

When observing polarised radio emission, three key phenomena provide valuable information about magnetic fields:

\begin{itemize}
    \item \textbf{Intrinsic Linearly Polarised Emission:} This refers to the inherent angle and amplitude of the linearly polarised radiation emitted by a source, offering direct insights into the magnetic fields within the emitting region itself.
    \item \textbf{Faraday Rotation:} Quantified by rotation measure (RM), Faraday rotation describes the rotation of the plane of polarisation as radio waves propagate through magnetised, ionised gas along the line of sight. This effect encodes information about the line-of-sight strength of magnetic fields in the intervening medium.
    \item \textbf{Faraday Complexity:} This encompasses the more intricate, frequency-dependent behaviours observed in polarisation spectra, revealing detailed structures and statistical variations in magneto-ionic environments both within and between sources.
\end{itemize}

These observational phenomena can be effectively interpreted using techniques such as rotation measure synthesis \citep{Burn1966,2005A&A...441.1217B,2009A&A...503..409H,2023MNRAS.522.1464R} and spectro-polarimetric modelling \citep[e.g.,][]{OSullivan2012,Anderson2016b,Sun2015}. For a comprehensive theoretical background on these methods and their applications, see \citet{2021MNRAS.507.4968F}.

While radio polarimetry has been used since the 1960s to study cosmic magnetism \citep{Westerhout1962, Wielebinski1962, CooperPrice1962, Slysh1965, Burn1966, GW1966}, progress was initially slow because each wavelength for each source often required a separate observation. Nevertheless approximately 500 RMs of extragalactic sources were known by the 1970s \citep{1979Natur.279..115S}, revealing coherent RM structures across the sky, even at high Galactic latitudes. Simultaneously, multi-frequency spectro-polarimetric behaviour was being leveraged to understand the intrinsic magneto-ionic structure of quasars \citep{Conway1974}, laying the foundation for modern broadband spectro-polarimetric radio surveys.

In the past two decades, there has been increasing interest in designing all-sky surveys to incorporate best-practice radio polarimetric techniques, leading to major improvements in yield and reliability over earlier surveys. A key driver of this evolution is the concept of an `RM grid' \citep{Beck2004}, which comprises a comprehensive catalogue of polarised background radio galaxies mapped across large, contiguous regions of the sky. RM grids serve as highly sensitive tools for probing magnetic fields in various astrophysical environments \citep[e.g.,][]{Gaensler2004,Feretti2004}, offering an ensemble of point-like probes that illuminate foreground magneto-ionic structures, thus enabling detailed studies of cosmic magnetism.

The power of RM grids is primarily determined by their sky coverage, areal density, precision, and reliability of the RM measurements. The first modern RM grids covered a few hundred deg$^2$ with a sky density of approximately 1 RM~deg$^{-2}$ \citep{2001ApJ...549..959G,2005Sci...307.1610G,2001ApJ...563L..31B,2003ApJS..145..213B,2007ApJ...663..258B}, driving significant advances in mapping coherent and random magnetic field components in defined regions, such as segments of the Galactic Plane or the Magellanic Clouds. A major advance in sky coverage was achieved by the polarisation re-analysis of the NRAO VLA Sky Survey by \citet{2009ApJ...702.1230T}, whose catalogue covered 85\% of the sky at a sky density of just over 1 RM~deg$^{-2}$. However, all of these RM grids suffered from small fractional bandwidths and a limited number of frequency channels, affecting the precision and reliability of the measurements. Subsequent programs, reviewed in \S\ref{sec_other_pol}, have improved on various aspects of this earlier work---often dramatically so---but the NVSS RM catalogue remains by far the largest single contribution to the set of known RMs \citep[see][]{VanEck2023}.

The POSSUM Collaboration\footnote{POSSUM is an open collaboration that welcomes applications from qualified astronomers and students. See \url{https://possum-survey.org/}.} (Polarisation Sky Survey of the Universe's Magnetism) is working towards a major leap forward in our understanding of the polarised and magnetised universe by delivering two major surveys leveraging the Australian SKA Pathfinder (ASKAP) radio telescope \citep{2021PASA...38....9H}. The first, SPICE-RACS (Spectra and Polarisation in Cutouts of Extragalactic Sources from the Rapid ASKAP Continuum Survey; \citealp{2023PASA...40...40T}, Thomson \emph{in prep.}), is a joint project with the ASKAP Observatory\footnote{SPICE-RACS is a sub-component of the ASKAP Observatory's official RACS Observatory-Led Project. It entails key contributions from both the POSSUM Collaboration and the ASKAP Observatory. SPICE-RACS is considered a POSSUM-led and -delivered joint science project executed in close collaboration with the RACS team and the Observatory, by agreement.}. It will cover an expansive 36,100\,deg$^2$ (87.5\% of the sky at $\delta < +49^\circ$), aiming to deliver an RM grid with a sky density of approximately 7 RMs deg$^{-2}$ in its initial all-sky data release, potentially doubling in future releases, at $\sim20''$ resolution and an RMS sensitivity of about 100 $\mu$Jy\,beam$^{-1}$. The second, the POSSUM survey itself \citep{Gaensler2010}, is planned to cover 20,630\,deg$^2$ (50.0\% of the sky) at greater depth and with a higher RM grid density, producing a comprehensive set of polarimetric data products with the best possible imaging performance from long-track synthesis observations. This paper provides a detailed description of POSSUM.

POSSUM has been developed over the past 15 years as a modern successor to the NVSS RM grid  \citep{2009ApJ...702.1230T} and polarisation maps. It aims to conduct a sensitive (18 \(\mu\)Jy\,beam\(^{-1}\) in Stokes \(I\), \(Q\), and \(U\); \citealp{2024AJ....167..226V}), high-resolution (20 arcsecond), full-polarisation sky survey over 800 to 1088\,MHz, covering 20,630\,deg$^2$ with a 30\% fractional bandwidth. POSSUM observes commensally with the Evolutionary Map of the Universe (EMU; \citealp{2011PASA...28..215N}, Hopkins et al., 2025, {\it in press}) survey, which focuses on Stokes \(I\) imaging over the same region and aims to create the most sensitive radio continuum survey of the Southern sky. EMU and POSSUM were designed in close consultation to ensure their outputs and catalogues complement one another (see Section \ref{1d_pipeline}). Additionally, the main survey will be supplemented by observations covering 15,470\,deg$^2$ (37.5\% of the sky) over 1296--1440\,MHz, leveraging commensal observations with the Widefield ASKAP L-band Legacy All-sky Blind Survey (WALLABY; \citealp{2020Ap&SS.365..118K}), which focuses on detecting neutral hydrogen (H\,\textsc{i}) in the nearby Universe.

One of POSSUM's core scientific products will be a dense, wide-area RM grid comprising up to one million linearly polarised background radio galaxies. This grid, with high-quality Faraday rotation information (median uncertainty in RM, $\delta$RM $\approx$ 1~rad\,m$^{-2}$), will enable cosmic magnetism to be probed across a broad spectrum of astrophysical phenomena. POSSUM's RM grid will be the most powerful ever obtained, thanks to four key features: a roughly thirty-fold increase in the sky density of polarised sources compared to previously available large-area surveys, an order-of-magnitude reduction in RM uncertainties, the availability of broadband spectro-polarimetric data ensuring precise and accurate RMs, and comprehensive coverage of the southern sky --- a region that is relatively underexplored using these techniques but which has become a focal point for many new and powerful telescopes that are ground-based in the Southern hemisphere and provide coverage across the electromagnetic spectrum.

POSSUM will also provide maps of polarised emission across the entire survey region, which will serve as an additional and complementary probe of the southern sky, revealing the polarisation properties of synchrotron emission, magnetic fields, and depolarising thermal gas in compact and extended structures such as radio galaxies, supernova remnants, galaxy clusters, nearby galaxies, and the Milky Way's interstellar medium (ISM).

POSSUM will revolutionise our understanding of magnetic fields in the ISM, in the large-scale Milky Way and halo, in other galaxies dominated by star formation, active galactic nuclei (AGN) or merger activity, inside and around galaxy groups and clusters, and in the overall cosmic web and intergalactic medium (IGM). In all these different locales, the magnetic field properties are either poorly mapped or completely unknown, and we thus lack an understanding of the role of magnetism in fundamental physical processes involving energy transfer, gas dynamics, star formation, and galaxy evolution. Underpinning all this are two deeper questions that POSSUM also plans to address: How were the first magnetic fields generated? And what processes have sustained, organised and strengthened these fields through to the present day? Enormous progress on all these issues will be enabled by the first sensitive large-area survey for polarisation and Faraday rotation that POSSUM represents.

The objectives of this paper are to outline the broad range of POSSUM's science objectives, detail the survey's capabilities, and place POSSUM in the context of contemporary and future polarisation surveys.  In \S\ref{sec_sci}, we describe POSSUM's primary scientific objectives  and outline a few example experiments. \S\ref{sec_obs} details the observational strategies that will generate the POSSUM dataset. In \S\ref{sec_processing}, we describe the various stages of data processing and the corresponding pipelines that will be employed. \S\ref{sec_products} provides an overview of the different POSSUM data products that will be made publicly available. \S\ref{sec_examples} presents illustrative examples of POSSUM data, showcasing the types of measurements and products that the survey will generate. In \S\ref{sec_others}, we explore synergies between POSSUM and other major ongoing surveys. Finally, \S\ref{sec:conclusion} summarises the expected contribution of POSSUM and its legacy in the era of the prospective SKA Cosmic Magnetism Key Science Project. Subsequent papers will provide detailed descriptions of individual POSSUM data releases and the technical nuances of data processing.

\section{SCIENCE DRIVERS}
\label{sec_sci}

POSSUM was conceived to harness the unique capabilities of the ASKAP telescope, delivering a state-of-the-art radio polarisation survey with a compelling combination of sky coverage, RM-grid density and precision, surface brightness sensitivity, image fidelity, and polarisation purity (see Section~\ref{sec_obs}). This design supports the robust core science program outlined in this section, while also maximizing serendipitous discovery potential and legacy value. Where relevant, we highlight synergies with other major ASKAP projects—EMU \citep[][Hopkins et al., 2025, {\it in press}]{2011PASA...28..215N}, WALLABY \citep{2020Ap&SS.365..118K}, VAST \citep{2013PASA...30....6M}, and FLASH \citep{Allison2022}—as well as with other large-area polarisation surveys such as LoTSS \citep{2017A&A...598A.104S} and VLASS \citep{Lacy2020}, and expand on these in \S\ref{sec_others}. 

\subsection{Intergalactic Magnetism}
\label{sec:intergalactic}

We do not know the origin of magnetism in the Universe. Broadly speaking there are two possible scenarios \citep{2002RvMP...74..775W,2006AN....327..395R}: a ``top-down'' model in which magnetism was generated everywhere in the Universe through plasma instabilities or phase transitions at very early times, or a ``bottom-up'' model in which magnetism was originally generated in stellar interiors and accretion disks, and then spread by various astrophysical processes. The key to distinguishing between these models are the magnetic properties of intergalactic gas since the initial
conditions are expected to be quickly washed out  by turbulent amplification processes in galaxy clusters \citep{Stasyszyn2010,Hackstein2019}. 

POSSUM offers significant advancements in measuring these properties. With its dense, wide-area RM grid, POSSUM is sensitive to magnetic structures on scales of megaparsecs, allowing us to probe magnetic field strengths down to the nanogauss level \citep{2014ApJ...790..123A}. Just as fast radio bursts (FRBs) have begun to reveal the gas distribution of the intergalactic medium (IGM) through dispersion measures \citep[DMs;][]{2020Natur.581..391M}, POSSUM's RM grid can be used to reveal the magnetic field of the IGM.

Specifically, RM~$\propto \int n_e B_\parallel dl$ (where $B_\parallel$ is the line-of-sight component of the magnetic field $\mathbf{B}$) can be combined\footnote{For experiments requiring only the direction of $B_\parallel$, RM alone suffices.} with FRB DMs, emission measures, and other tracers to directly estimate $B_\parallel$, which can then be used to infer the total magnetic field strength under certain assumptions. Additionally, RMs are exceptionally sensitive probes of $n_e$ when other tracers are unavailable, allowing magnetic fields to ``illuminate" the ionised gas via Faraday rotation. The effectiveness of this approach was demonstrated in POSSUM's first Early Science paper, which mapped the gaseous extent of the Fornax cluster purely through an RM grid \citep{Anderson2021}.

\subsubsection{The Intergalactic Medium and Cosmic Web}

Approximately 30--50\% of the warm-hot intergalactic medium (WHIM) is believed to reside in intergalactic filaments, which play a pivotal role in galaxy evolution and large-scale gas flows \citep{CO1999,Shull2012}. Yet direct observations of the WHIM remain challenging \citep[e.g.,][]{Parimbelli2023,Zhang2024arXiv,DiMascolo2024arXiv,Medlock2024}, leaving its magnetic properties largely unexplored. These fields are crucial not only for understanding cosmic magnetism, but also for tracing the poorly probed physical conditions within the WHIM.

Recent RM-grid studies have constrained the intergalactic magnetic field (IGMF) in large-scale structure \citep{2019ApJ...878...92V,2020MNRAS.495.2607O,2020A&A...638A..48S,2021MNRAS.503.2913A}, culminating in a detection based on \(\sim1000\) LOFAR RM sources \citep{2022MNRAS.512..945C,2022MNRAS.515..256P} and ongoing refinements \citep{Carretti2023,2025A&A...693A.208C}. Early POSSUM data likewise revealed denser cosmic web phases connected to galaxy groups \citep{Anderson2024arXiv}, highlighting the complementary advantages of low-frequency and GHz-band RM surveys.

Over large sky areas at low redshifts, LoTSS’s \(\sim10\times\) lower per-source RM error and POSSUM’s \(\sim100\times\) higher RM density (see \S\,\ref{sec_survey_yield}) could be thought to yield comparable statistical power, since the ensemble RM uncertainty typically scales as \(N^{-1/2}\) (\(N\) is the total number of RM measurements). However, as shown in \S\,\ref{subsec:rmgrid}, the typical angular scales and magnitude of RM variations are in practice better recovered with higher RM density. Furthermore, when combined with POSSUM’s higher observing frequency and redshift reach \citep[extending to \(z\sim2.5\) vs.\ \(z\sim1.5\);][]{Bonaldi2019,Carretti2022,Piccirilli2023}, this enhanced RM grid enables the survey to:
\begin{enumerate}
    \item ``resolve'' cosmic web filaments of width $\sim6\,\mathrm{Mpc}$ (with multiple RMs across each filament) out to $z\sim2.5$, compared to $z\sim0.1$ for LoTSS or NVSS,
    \item probe denser cosmic-web junctions (e.g.\ with groups or cluster outskirts; \citealp{Anderson2021,Anderson2024,Loi2025}) that may substantially depolarise at lower frequencies \citep[see][also Section \ref{sec_survey_design}]{2025A&A...693A.208C}, and
    \item disentangle cosmic filament RMs from the Galactic ISM foreground via superior foreground reconstructions on sub-degree scales.
\end{enumerate}

Thus, POSSUM’s large, dense RM grid can measure the IGMF over a broader swath of cosmic environments and epochs, particularly when paired with redshift data (see Hopkins et al., 2025, {\it in press}). This opens avenues for moving beyond purely statistical RM-stacking approaches to more nuanced analyses---e.g., applying RM structure-function techniques to small ensembles of filaments, or mapping the RM patterns of individual nearby filaments \citep[e.g.][]{Loi2025} or superclusters \citep[e.g.][]{Alonso2025_inprep,2025arXiv250308765P}. By furnishing a more detailed view of the magnetised cosmic web and its junctions with groups and clusters---and in synergy with low-frequency RM grids---POSSUM will provide key insights into the seeding and amplification of cosmic magnetic fields \citep[e.g.,][]{2016RPPh...79g6901S,2017CQGra..34w4001V}.

\subsubsection{Galaxy Clusters and Groups}

Magnetic fields play a crucial yet still poorly understood role in the evolution of galaxy clusters and groups, particularly within their intracluster medium (ICM) and intragroup medium (IGrM). These fields are essential for accelerating and guiding cosmic rays \citep[e.g.,][]{RP2023} and for regulating mixing and thermal conduction \citep[e.g.,][]{Barnes2019}. In rich clusters, the ICM typically hosts fields of $0.1$--$10\,\mu$G \citep[][]{2018SSRv..214..122D}, likely amplified from primordial seeds whose exact origin remains unclear.

\begin{figure*}[ht]
    \centering
    \includegraphics[width=\textwidth]{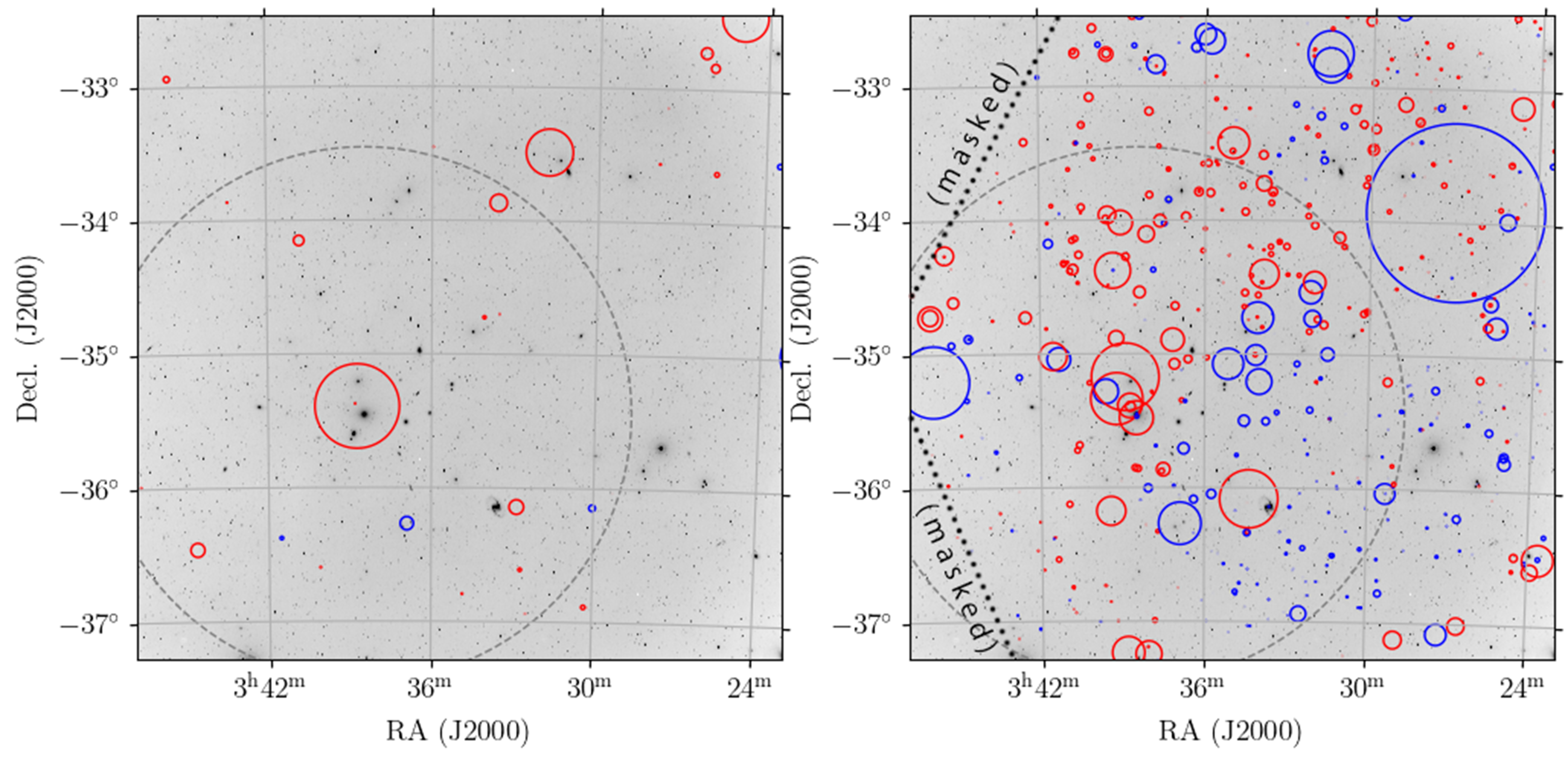}
    \caption{Comparison of NVSS \citep[left;][]{2009ApJ...702.1230T} and POSSUM (right) Band~1 RMs within the Fornax cluster’s virial radius (grey dashed circle), showcasing POSSUM's transformative capability to probe magnetised gas in clusters, groups, and many other $\sim$degree-scale extragalactic objects. The background is a Digitized Sky Survey optical image (greyscale). Solid circles indicate the position, magnitude, and sign of RMs measured by each survey, with diameters proportional to (RM)$^2$ and colours indicating RM sign (red for positive, blue for negative). While only a few clusters or groups contain $\gtrsim10$ NVSS RMs at sub-virial radii, POSSUM’s RM grid will sample thousands of these systems, including over a dozen with 100 or more RMs.}
    \label{fig:fornax_RMs_comparo}
\end{figure*}

The work of \citet{Anderson2021} demonstrated POSSUM’s transformational capability for studying the magnetised ICM in detail. As shown in Figure~\ref{fig:fornax_RMs_comparo}, ASKAP commissioning data on the Fornax cluster achieved an RM grid density of 27\,RMs\,deg$^{-2}$, revealing multiple cluster components and tracing the ICM out to the virial radius—well beyond its apparent X-ray extent. Building on this, the full POSSUM survey will map several other nearby clusters (e.g.\ Coma, Shapley, Virgo), each with \emph{hundreds} of RM measurements inside the projected virial radius, compared to the $\lesssim10$ previously available \citep[e.g.,][]{2010A&A...513A..30B,2013MNRAS.433.3208B,2021MNRAS.502.2518S}. At least 100 additional high-mass, low-$z$ clusters will have $\gtrsim50$ RM sight lines, filling a major observational gap. Such a substantial increase in RM statistics will facilitate Faraday rotation analyses of large-scale ICM fields and depolarisation studies of smaller-scale features \citep[][]{2010A&A...513A..30B}, including how the field strength may scale with gas density \citep[e.g.][]{Osinga2024arxiv}. Meanwhile, thousands of more distant or smaller clusters will still each have on the order of 1--10 POSSUM RM measurements, making RM stacking experiments feasible in bins of redshift, cluster richness, environment, and other key formation-related parameters \citep{2009MNRAS.400..646K}. 

In addition, the IGrM of galaxy groups---the most common environment for galaxy evolution---are even less thoroughly characterised. Observations suggest the IGrM is shaped by shallower gravitational potentials, mergers, outflows, and galactic motions \citep{Lovisari2021,Oppenheimer2021}, yet direct RM data for these lower-mass halos have been scarce. POSSUM’s order-of-magnitude increase in RM density now extends RM-grid techniques to hundreds of group-scale systems, revealing magnetised gas that can extend well beyond group boundaries into the cosmic web \citep{Anderson2024arXiv}.

A key strength of POSSUM’s wide sky coverage is that it also samples the outskirts of clusters/groups and the Galactic ISM foreground, regions typically undersampled by targeted RM campaigns. This comprehensive approach supports both single-object studies and ensemble statistical methods. Furthermore, clusters and groups host polarised radio galaxies and relics that trace field configurations via 2D maps of RM, intrinsic polarisation, or depolarisation \citep[e.g.,][]{2019SSRv..215...16V,2019MNRAS.489.3905S,2022A&A...666A...8S}. Capturing such low surface brightness ($\lesssim1\,\mu\mathrm{Jy\,arcsec}^{-2}$) emission on $\lesssim10'$ scales requires a sensitive survey at sub-$1\,\mathrm{GHz}$ frequencies, such as POSSUM, which can detect RM variations at the $\sim10\,\mathrm{rad\,m}^{-2}$ level from shocks or cluster interactions \citep[e.g.,][]{2013MNRAS.433.3208B}.

Finally, these observations will feed back into numerical simulations of the ICM/IGrM, whose hot ($10^{7}$–$10^{8}$ K), dilute ($10^{-2}$–$10^{-3}$ cm$^{-3}$), and weakly collisional nature \citep{2022hxga.book...56K} results in complex, poorly understood coupling across a wide range of scales, posing significant challenges. While simulations have become increasingly powerful \citep{2024MNRAS.528..937A}, they still lack comprehensive observational constraints. By providing a dense RM grid across a wide range of cluster/group masses and redshifts, POSSUM will supply the necessary observational data to anchor these models. This will not only illuminate how magnetic fields evolve in weakly collisional plasmas over cosmic time but also clarify how these fields differ from those in collisional plasmas, where MHD is typically applicable.

\subsection{The Magnetism of Galaxies}
\label{sec:galactic}

Magnetic fields are crucial to the dynamics and evolution of galaxies, influencing the energy balance of the ISM, regulating star formation, and controlling gas flows into and out of galactic disks. While detailed observations have revealed typical magnetic field strengths and structures \citep[][updated \citeyear{arXiv:1302.5663v11}]{2013pss5.book..641B}, it remains unclear whether these fields are shaped by standard dynamo processes or faster, more complex mechanisms. By simultaneously providing polarisation spectra of background radio sources \emph{and} ISM emission, POSSUM will probe magnetic field structures across a wide range of scales, providing insights into the complex 3D magneto-ionic structures of the Milky Way and other galaxies \citep[e.g.,][]{2018MNRAS.477.2528B, 2021MNRAS.508.1371S}.

\subsubsection{The Large-Scale Milky Way}

Quantifying the strength and structure of the Milky Way's magnetic fields in detail is crucial for understanding their role in various stages of the Galactic gas cycle, testing the amplification and ordering mechanisms of Galactic magnetic fields, and tracing the astrophysical conditions of our Galaxy back to its infancy. POSSUM will enable the creation of an all-sky interpolated map of Galactic Faraday rotation, improving upon previous efforts by \citet{2012A&A...542A..93O} and \citet{2022A&A...657A..43H} with far superior precision and angular resolution. This map will reveal the magnetic fields of our Galaxy across an enormous range of scales and allow for rigorous testing of the $\alpha$--$\Omega$ dynamo theory for the magnetism of the Galactic disk. As demonstrated by several studies \citep[e.g.,][]{2011ApJ...728...97V,shanahan2019,2020MNRAS.497.3097M}, a dense RM grid is essential for accurately recovering large-scale magnetic structures in the Milky Way. Additionally, POSSUM's RM grid will probe the halo field at intermediate Galactic latitudes, enabling us to reconstruct the halo magnetic field geometry \citep[e.g.,][]{2014A&A...561A.100F,2019A&A...623A.113S,2022ApJ...940...75D} and determine whether its origin lies in a halo dynamo, disk-driven outflows, or the Parker instability \citep{2013ApJ...764...81M}.

\subsubsection{Milky Way Interstellar Medium}
\label{sec_ISM}
Any understanding of the ISM requires an understanding of its magnetic properties. POSSUM will offer a transformative view of two magnetised ISM components: individual interstellar objects such as supernova remnants, \HII\ regions, molecular clouds, and wind bubbles; and the turbulent cascade that transfers global energy injection to small scales where it dissipates as heat. Combined with data from ASKAP's GASKAP-H\textsc{I} survey \citep{Dickey2013}, POSSUM will provide an unparalleled perspective on the interplay between the ISM and magnetic fields.

\paragraph{Magnetised Turbulence in the ISM:} POSSUM's RM grid will provide a complete census of the magnetised turbulent ISM as a function of the environment, on scales from degrees down to $<10''$ (using Faraday complexity to model small-scale fluctuations, e.g., \citealt{Anderson2015,2021MNRAS.502.3814L,2023MNRAS.518..919S,2024A&A...686A.104R,2024AJ....167..226V}).  Data over such a large inertial range will erase the long-standing artificial separation between ``small-scale'' vs ``large-scale'' magnetic fields in the ISM \citep[e.g.,][]{2015ASSL..407..483H}, finally providing a single consistent picture. Probing magnetic fields down to arcsecond scales will allow us to study the fine-scale structure of ISM magnetic fields \citep{2020PhRvF...5d3702S, 2021MNRAS.502.2220S} and may provide insights into small-scale turbulent dynamo theories \citep{2019JPlPh..85d2001R, 2021PhRvF...6j3701S}. Interpreting these observations will require careful comparison with realistic 3-D magnetohydrodynamic simulations which capture turbulence driven by supernova explosions and the effect of local star-forming \HII\ regions \citep[e.g.,~][]{2013MNRAS.432.1396G, 2018MNRAS.480.3916M, 2021ApJ...910L..15G}, as well as sophisticated analysis of observational effects like diverging lines of sights \citep[e.g.,~as discussed in][]{2009A&A...495..697W, 2019ApJ...885...15R}. Comparing POSSUM RMs with a synthetic RM grid produced from such simulations will allow a direct comparison of observations with existing theories and simulations of turbulent dynamo theory.

Furthermore, POSSUM's pilot observations have demonstrated excellent sensitivity to diffuse polarised emission from the ISM, which on their own \citep{2011Natur.478..214G} and when merged with single-dish polarimetry from the Parkes PEGASUS survey (\S\ref{sec_single_dish_surveys}), can provide direct `Faraday tomography' of the 3D magnetised multi-phase ISM \citep{2017A&A...597A..98V,2019MNRAS.487.4751T}. 

\paragraph{Individual Galactic Objects:} Polarisation data from individual interstellar objects such as supernova remnants or H\,\textsc{ii} regions provide {\em in situ}\ probes of the Galactic magnetic field and of cosmic ray acceleration mechanisms \citep{2011ApJ...736...83H,2017A&A...597A.121W,2021RNAAS...5...12S}. RM grids have been used to study just a handful of large objects \citep[e.g.,][]{2011ApJ...736...83H,2018ApJ...865...65C,2018A&A...614A.100T,2021MNRAS.508.3921J}, which is insufficient to understand trends with evolutionary stage or with location in the Galaxy. POSSUM's dense RM grid together with EMU's Stokes $I$ images will hugely expand the number of such sources that we can study. In addition, many synchrotron sources themselves will emit diffuse linear polarisation (synchrotron sources), while thermal sources will depolarise diffuse Galactic background polarisation. These effects can be combined with the RM grid to understand the 3D intrinsic field geometry and thermal gas properties of the ISM \citep{2001ApJ...549..959G,2010ApJ...712.1157H}, along with the thermal gas-magnetic field connection \citep{2022MNRAS.514..957S}.

\subsubsection{The Magellanic Clouds and Other Galaxies}
\label{sec:magellanic}
POSSUM's RM grid will reveal the magnetic field structures of the Magellanic Clouds from large to small scales \citep{2024MNRAS.534.2938J}, providing us with a detailed view of galaxies' magnetism at a known distance.  Together with the GASKAP-HI survey, POSSUM will resolve the turbulent cascade of the magnetoionic medium across a wide range of astrophysical environments distinct from those of our own Milky Way. 

POSSUM's dense RM grid and diffuse polarisation imaging  will recover the regular magnetic field structure in resolved galaxies in the nearby Universe. Sensitive measurements of diffuse polarised emission with high RM precision towards $\sim 200$ nearby resolved galaxies will  map the structure of large-scale fields in a range of different galaxy types.  There are also of order ten large, highly resolved galaxies in the survey region for which POSSUM will have enough background RMs to measure the structure of large scale magnetic fields.   When RM grid information is coupled with polarised diffuse emission it is particularly powerful for revealing the 3-D structure of a galaxy  \citep[e.g.][]{2012ApJ...759...25M,2020A&A...642A.118K}.

The POSSUM RM grid density will also illuminate the circumgalactic environment and disk-halo interface of dozens of other galaxies from the WALLABY survey across a range of environments, through measurements of diffuse polarisation and associated RMs \cite[see][for an example of what is currently possible statistically with sparser RM grids]{2023A&A...670L..23H}. 

A single gas-rich galaxy can dominate the RM on a line of sight, despite being too faint to detect directly in line or continuum radio emission. Such objects have been detected as depolarisation silhouettes against extended radio lobes \citep[e.g.,][]{fomalont1989depolarization}, and by correlation with intervening absorbers found in optical/UV spectra \citep{Kronberg1990,2014ApJ...795...63F}. Very few silhouettes will be resolved by POSSUM, but it will be possible to exploit known Mg\,{\sc ii} and damped Ly$\alpha$ absorbers that fall in the POSSUM footprint \citep[noting that most current catalogues are strongly biased to the northern hemisphere; e.g.,][]{Zhu2013}. Moreover, Faraday rotation and depolarisation  data on intervening \HI\ absorption systems seen by ASKAP's FLASH survey will provide information on the magnetisation and ionisation state of large numbers of galactic disks at $0.4 <z<1.0$. Intervening \HI\ lines are found at impact parameters $<20\ \mathrm{kpc}$ \citep{Gupta2010} enabling us to directly study galactic disks. By assuming an effective ionization fraction, the 21-cm absorption column can be used to estimate the magnetic field strength in the absorbing medium. 

Most studies of ISM magnetic fields focus on star-forming galaxies \citep{2015A&ARv..24....4B}, but probing magnetic fields in quiescent (non-star-forming and elliptical) galaxies is also crucial for testing theories of turbulent dynamos \citep{1996MNRAS.279..229M, 2021ApJ...907....2S}. Direct observation of these magnetic fields is challenging due to the lack of significant synchrotron emission in quiescent galaxies and the potential contamination from central AGNs. However, POSSUM's dense RM grid provides a unique opportunity to investigate these fields. By cross-matching background RMs with catalogues of quiescent galaxies \citep[see, e.g.,][]{2021MNRAS.508.1371S}, we can detect excess Faraday rotation and gain insights into the magnetic field properties of these galaxies, advancing our understanding of turbulent dynamos.

\subsubsection{Active Galactic Nuclei}
\label{sec:AGN}

Broadband polarisation studies have inferred the structure of \emph{spatially unresolved} polarized radio jets from AGN by analysing spectropolarimetric interference effects generated by jet structure \citep{OSullivan2012,Anderson2016,Kim2016,2017MNRAS.469.4034O,Pasetto2018} and by examining the time-evolution of these effects \citep{2019MNRAS.485.3600A,2019MNRAS.487.3454M}. These observations provide valuable insights into the physics of nuclear outbursts and the mechanisms involved in the launching, confinement, and bulk transport of AGN jets.  However, the studies conducted so far have only encompassed a few hundred sources. POSSUM will provide polarisation and RM data for approximately $10^5$~unresolved blazars, significantly enhancing our understanding of the RM properties of relativistic jets close to their launching sites.

An especially promising area of study involves the intrinsic magnetoionic structure of AGN using the \HI\ absorber sources identified by FLASH. Absorption by nuclear discs has been found in front of the counter-jets in large-scale twin-jet sources, but never the approaching jets, as expected from AGN models \citep{Morganti2018}. Such data can probe the nature and dynamics of the gas feeding the central black hole, specifically its magnetisation and ionisation fraction, which are key variables for the infall of material and the formation of AGN jets. 

High detection rates of absorbers have been found in young, compact, recently-activated radio AGN, whose jets and lobes are embedded within the ISM of the host galaxy \citep{Gupta2006}.   Nearby examples show patchy absorption on scales from tens to thousands of parsec, in part because only one lobe is likely to be behind the gas disc. The radio jets are driving outflows in the ISM, and molecular, atomic, ionized, and in at least one case, shock-heated X-ray emitting gas are intimately associated. In the few cases where polarisation has been resolved, a wide range of RMs are found in such material, leading to patchy depolarisation \citep[e.g.,][]{Floyd2006,Hardcastle2012}. In the FLASH redshift range ($0.4< z < 1$), associated absorbers will be unresolved by POSSUM, so statistical analysis will be required to define the relation between Faraday depth and \HI\ absorption and to identify the cosmological evolution of that relationship. This work will also identify targets for VLBI observations to resolve both \HI\ absorption and polarisation.

\subsection{The Interplay Between Galactic and Intergalactic Magnetism}
\label{sec:interplay}

Magnetism plays a crucial role in both influencing and being influenced by the gaseous and energetic interactions between galaxies and the circumgalactic medium (CGM). This includes processes such as star formation \citep{2020arXiv201212905S}, AGN-driven outflows \citep{2021A&A...653A..23V}, and gas accretion \citep{2018ApJ...865...64G}. Understanding these processes requires answers to questions such as: How far do galactic magnetic fields extend into the CGM?  How strong is the magnetism in halos and the CGM? How does this depend on environment, and how has this evolved over cosmic time? POSSUM will address these issues for the Milky Way and other star-forming galaxies, for interacting galaxies, and for AGN-driven outflows.

\subsubsection{Galaxy Halos and the Circumgalactic Medium}
\label{sec:halos}

POSSUM's RM grid will allow us to measure the strength and configuration of magnetic fields in regions of gas accretion and fountain mixing in the Milky Way's halo, in nearby galaxies, and in more distant galaxies projected against background quasars.

An all-sky grid with $\gtrsim 30$\,RM\,deg$^{-2}$ and a precision of $\delta$RM~$\lesssim 5$--10\,rad\,m$^{-2}$ is required to analyse magnetic fields in relevant structures such as high-velocity clouds \citep[e.g.,][]{2019ApJ...871..215B,2023MNRAS.526..836J}, and $\delta{\rm RM} \lesssim 1$\,rad\,m$^{-2}$ is needed to detect halo fields around external edge-on spirals, as demonstrated by recent results from MeerKAT \citep{2023A&A...678A..56B} and LOFAR \citep{2023A&A...670L..23H}. The higher the grid density, the more objects that can be studied, complementing and significantly expanding the sample of individual halo observations that have been the primary approach until now \citep[e.g.,][]{2020A&A...639A.112K}.

In addition, thousands of quasars show optical absorption lines in their spectra due to halos of intervening galaxies. Polarisation and RM observations of these sightlines can directly probe halo and CGM magnetic fields for a large statistical sample across cosmic time. However, current datasets are too small to establish clear conclusions \citep{2014ApJ...795...63F,2020ApJ...890..132M,2020MNRAS.496.3142L,2021MNRAS.508.1371S}.

The extensive and high-quality RM grid provided by POSSUM will enable major progress, allowing us to track halo magnetic fields as a function of redshift, equivalent width, impact parameter, and galaxy orientation. To derive meaningful statistical conclusions from this approach, precisions on individual measurements of $\delta{\rm RM} < 10$\,rad\,m$^{-2}$ and $\sim$hundreds of intervening galaxies are required. Estimating magnetic field properties in the CGM is crucial for determining the role of magnetic fields in the evolution of galaxies.

\subsubsection{Galaxy Interactions}

The few detailed studies of magnetism in interacting galaxies show that the total magnetic field strength is enhanced relative to that of isolated galaxies, and that highly ordered fields can be found in tidal features \citep{2011A&A...533A..22D}. Stretching and compression driven by galaxy interactions may therefore offer a mechanism to amplify large-scale ($\gtrsim1$~kpc) fields more rapidly than the $\sim10^9$~yr timescale of the $\alpha$-$\Omega$ dynamo and thus reconcile observations of early and nearby galaxies. Observations of diffuse polarisation and RMs in many more interacting systems are required to understand how interactions drive this process \citep{2017MNRAS.464.1003B}. Furthermore, just as POSSUM can use the RM-grid to find missing ionised gas in the cosmic web and galaxy clusters, it may also trace the ionised gaseous web induced by interactions between galaxies.

The Magellanic Clouds and Milky Way provide us with an especially unique observational perspective on galaxy interactions.  The Magellanic Bridge connecting the Clouds has a coherent magnetic field \citep{2017MNRAS.467.1776K}. POSSUM's RM grid will be able to tell us if the massive gaseous Magellanic Stream, which trails the Magellanic Clouds across $\gtrsim 100^\circ$ of the sky, could also host a significant magnetic field.  As a whole, the Magellanic system (which has a far larger angular extent than just the two Clouds) gives a special opportunity to disentangle the effects of tidal forces and ram pressure on magnetic field amplification.  If the Magellanic Stream has a $\approx 0.1$--$1\,\mu$G magnetic field \cite[as seen by][in the Magellanic Bridge]{2017MNRAS.467.1776K}, a POSSUM RM grid with a precision of $\delta$RM~$\lesssim$~5~rad~m$^{-2}$ is needed to separate its  small Faraday rotation signature from the Milky Way foreground.  

\subsubsection{Outflows from Active Galaxies}\label{sec:243}

Kinematic outflows from AGN have multiple, important impacts on the magnetised thermal plasma in which they are embedded: they inflate cavities, heat the ICM, redistribute metals, and regulate the central star formation rate through their modification of infalling, cool gas \citep{2019arXiv190109448R,2024PNAS..12102435C}. Faraday RMs and depolarisation can provide excellent tracers of these outflow-environment interactions \citep[e.g.,][]{Carilli1988,Feain2009,2011MNRAS.413.2525G,2012MNRAS.423.1335G,Anderson2018,Banfield2019,Knuettel2019,Anderson2022,Rudnick2024}.

POSSUM will break new ground on both statistical AGN studies and observations of well-resolved individual sources (see Fig.~\ref{fig:rgs}). Large new samples from POSSUM of fractional polarisation, depolarisation, and RM for AGNs will allow subsamples to be binned by luminosity, redshift, size, morphology, environment, and host galaxy properties. The goal of such studies will be to disentangle the role of AGN properties (e.g., black hole mass, spin, jet kinetic power) from environmental interactions (external density, pressure and its gradient, turbulence, etc) and from evolution with redshift. 

Differential RMs across and within radio galaxies are mainly generated by magnetic fields in their immediate environment, as demonstrated by systematic differences between the RM structures of approaching and receding lobes \citep{Garrington1988,2021MNRAS.508.1371S} and correlations with size \citep{Strom1973,Strom1988,2016ApJ...829....5L}. The redshift evolution of these structures can provide critical information on the growth of cosmic magnetic fields, \emph{if} it can be separated from luminosity, size, and morphology effects. 
Previous samples \citep[e.g.,][]{2012ApJ...761..144B} have given inconsistent results and were too small to separate redshift effects. POSSUM will allow clean separation of luminosity and $z$, and factors such as linear size and cluster environment, both by creating sub-samples and by exploiting POSSUM's large $\lambda^2$ coverage.

Complementing the absorption line studies described in \S\ref{sec:halos}, statistical studies of small double radio galaxies will allow an extensive characterisation of the magnetised CGM as a function of host magnitude, redshift, and environment. 
Intriguing banded RM patterns \citep{Bicknell1990,2011MNRAS.413.2525G,2012MNRAS.423.1335G,2019MNRAS.487.3432M,2020ApJ...903...36S} have been attributed to modification of the local fields by the jet interaction, but could also be illuminating organised structure in the cluster fields. The prevalence of such structures is still unknown. 
The many thousands of POSSUM sources with extents $\gtrsim 2'$ will increase the number of detailed RM maps by 1--2 orders of magnitude.   Statistical analysis of the RM structure in the best-resolved objects is the only practical way to measure the turbulent power spectrum of the surrounding medium, and comparison with magnetohydrodynamic simulations will help infer the underlying physics \citep[e.g.,][]{2014MNRAS.443.1482H}. Furthermore, we know that some entrainment of thermal gas does occur in AGN outflows \citep[e.g.][]{2013ApJ...764..162O,Anderson2018}. By tracing internal Faraday rotation via broadband polarimetry, POSSUM will allow a comprehensive study of this poorly probed phenomenon.

\subsection{Other Science Topics}
\label{sec:other}

POSSUM will enable a broad range of scientific investigations beyond those described in detail above. Some specific examples include:

\begin{itemize}
\item Using POSSUM's RM grid in conjunction with FRBs, both to correct for Galactic foregrounds of FRB RMs, and to combine with an FRB ``dispersion measure grid'' to study the Milky Way's halo \citep[][]{2022MNRAS.516.4739P};
\item Constraining properties of unusual objects, for instance Odd Radio Circles \citep{2021PASA...38....3N}, via their spectra and polarisation, in combination with EMU data;
\item Searches for violations of Lorentz invariance, quintessence, and axion-like particles via cosmic birefringence \citep[see][and references therein]{2021PhRvL.126s1102B}; and
\item Studying both linear and circular  polarisation properties of unusual transients, in conjunction with VAST data.

\end{itemize}

\section{OBSERVATIONS AND SURVEY DESCRIPTION}
\label{sec_obs}

\subsection{The Australian SKA Pathfinder (ASKAP)} \label{sec_ASKAP}

ASKAP is a radio interferometric array located at Inyarrimanha Ilgari Bundara, the CSIRO Murchison Radio-astronomy Observatory in outback Western Australia \citep[WA;][]{2021PASA...38....9H}. Situated within the Australian Radio-Quiet Zone WA, the array benefits from exceptional protection against terrestrial radio frequency interference (RFI). This is particularly advantageous for the critical 800--900 MHz range in POSSUM's main survey band, which is generally unusable at other synthesis arrays due to its allocation for mobile phone networks. At ASKAP, all such transmitters are far beyond the horizon, with interference only occurring during rare atmospheric ducting events. When these events do happen, they affect bands of about 10 MHz wide for just a few hours, and the data is automatically flagged, resulting in a negligible impact on POSSUM's data quality.

The ASKAP array comprises 36 fully steerable 12-meter Alt-Az mounted dishes equipped with an innovative third roll axis (described below), distributed over baselines ranging from 22 meters to 6 kilometres. The telescope can observe in the 700 to 1800 MHz frequency range. Each dish is equipped with a phased array feed (PAF), providing an instantaneous field of view of approximately 30 square degrees at 800 MHz, using 36 electronically formed dual linearly polarised beams. These beams are arranged to optimize both survey speed and uniformity of sensitivity across the field of view. 

For each formed beam, the ASKAP correlator generates full polarisation visibility spectra ($XX$, $XY$, $YX$, and $YY$) with 15,552 fine spectral channels across an instantaneous bandwidth of up to 288 MHz. These data are organized into \texttt{CASA} measurement sets by the ASKAP ingest system \citep{maxim2020} and are recorded on disk at the Pawsey Supercomputing Research Centre for reduction by the ASKAP Pipeline.

Each beam requires independent calibration. During the beamforming process, an on-dish calibration system at the vertex of each antenna radiates noise that is used to correct any relative phase differences between the $X$ and $Y$ formed beams of the antenna \citep{Chippendale2019}, ensuring high intrinsic Stokes-$V$ purity. While the noise sources are switched off during science observations to avoid residual pickup by nearby antennas, they are periodically reactivated to correct for slow drifts in the gains of the PAF elements, which could otherwise degrade beam quality.

The instrumental bandpass and on-axis instrumental polarisation calibration for each beam are typically derived shortly before or after the science observations, usually once every 24 hours, using the primary calibrator PKS~B1934--638. This calibrator is observed for a few minutes at the centre of each formed beam, and the resulting calibration is then applied to the corresponding science data during processing (see \S\ref{sec_processing} for details).

Models of the ASKAP primary beams are derived from holographic measurements using the unresolved, unpolarised calibrator PKS~B0408--65. These observations allow for the accurate modelling and correction of the off-axis instrumental polarisation response. A comprehensive characterisation of ASKAP's off-axis polarisation leakage performance for various use cases---such as band-averaged RM synthesis or frequency-dependent depolarisation modelling---will be provided with the first data release. Early results indicate that ASKAP's leakage from Stokes $I$ into $Q$, $U$, and $V$ is generally around 0.2\% across most of the observing footprint, averaged across the band (with a caveat discussed in \S\ref{sec:obsproducts}). 

This exceptional wide-field performance is made possible by ASKAP's axisymmetric antenna design and third roll axis, which maintain a constant sky orientation, akin to that of an equatorially mounted telescope. This capability is fully realized through routine automated observations and data processing, making ASKAP the only GHz-band radio telescope in the world capable of delivering high-purity polarisation data over wide fields almost immediately after observation. Additionally, ASKAP's distinctive observing mode---where each sky area is typically covered by multiple independent beams, and each formed beam observes multiple independent regions of the sky, with all data readily accessible in the online archive (\S\ref{sec_QA})---provides unique and powerful capabilities to identify and correct residual leakage or other biases during post-processing, using the statistics accumulated from field sources present in the main survey observations.

\begin{figure*}[h!]
\includegraphics[clip, trim=1.3cm 1.5cm 1.3cm 1.5cm, width=1.00\textwidth]{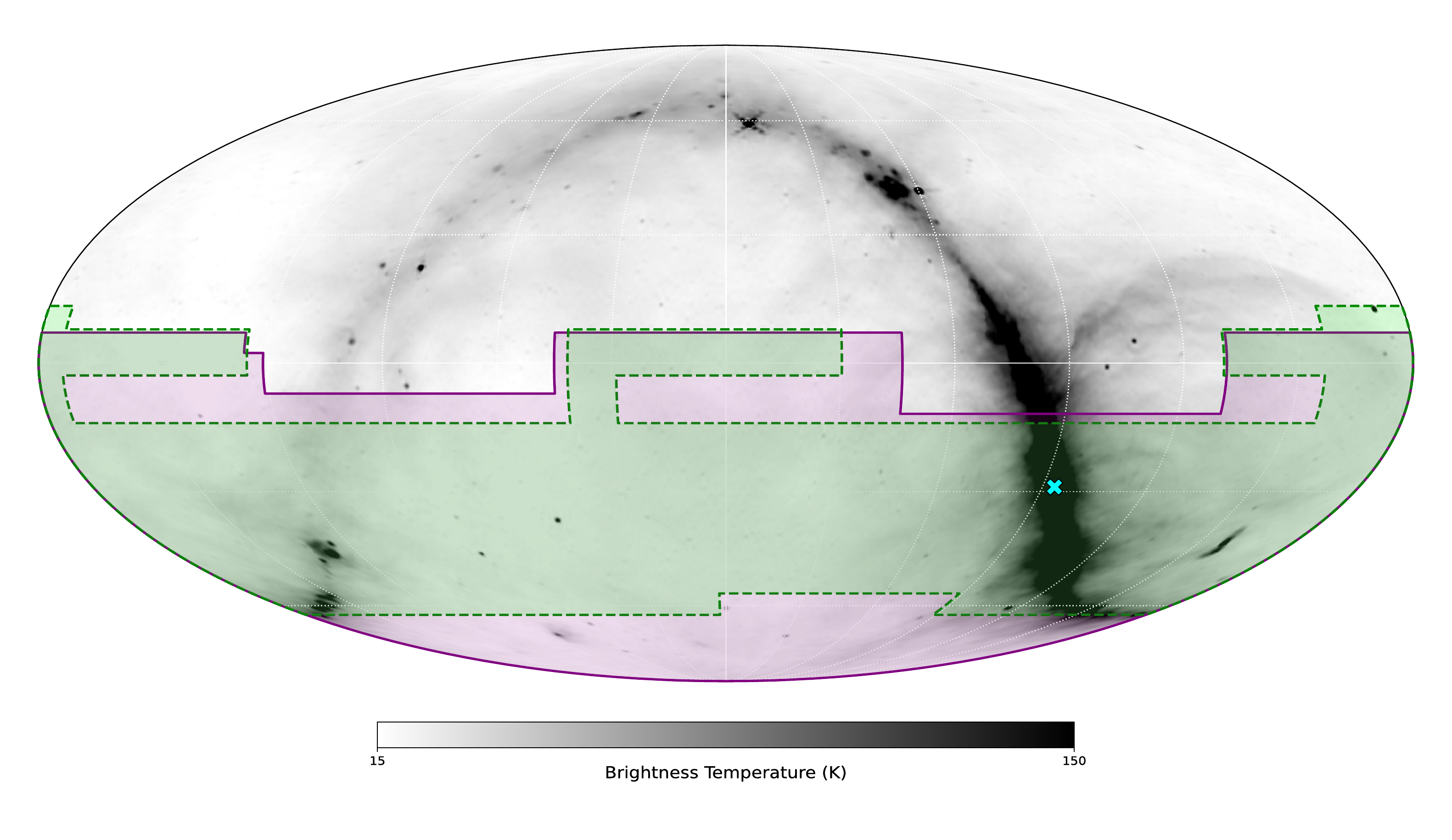}
\includegraphics[clip, trim=1.3cm 2.0cm 1.3cm 1.5cm, width=1.00\textwidth]{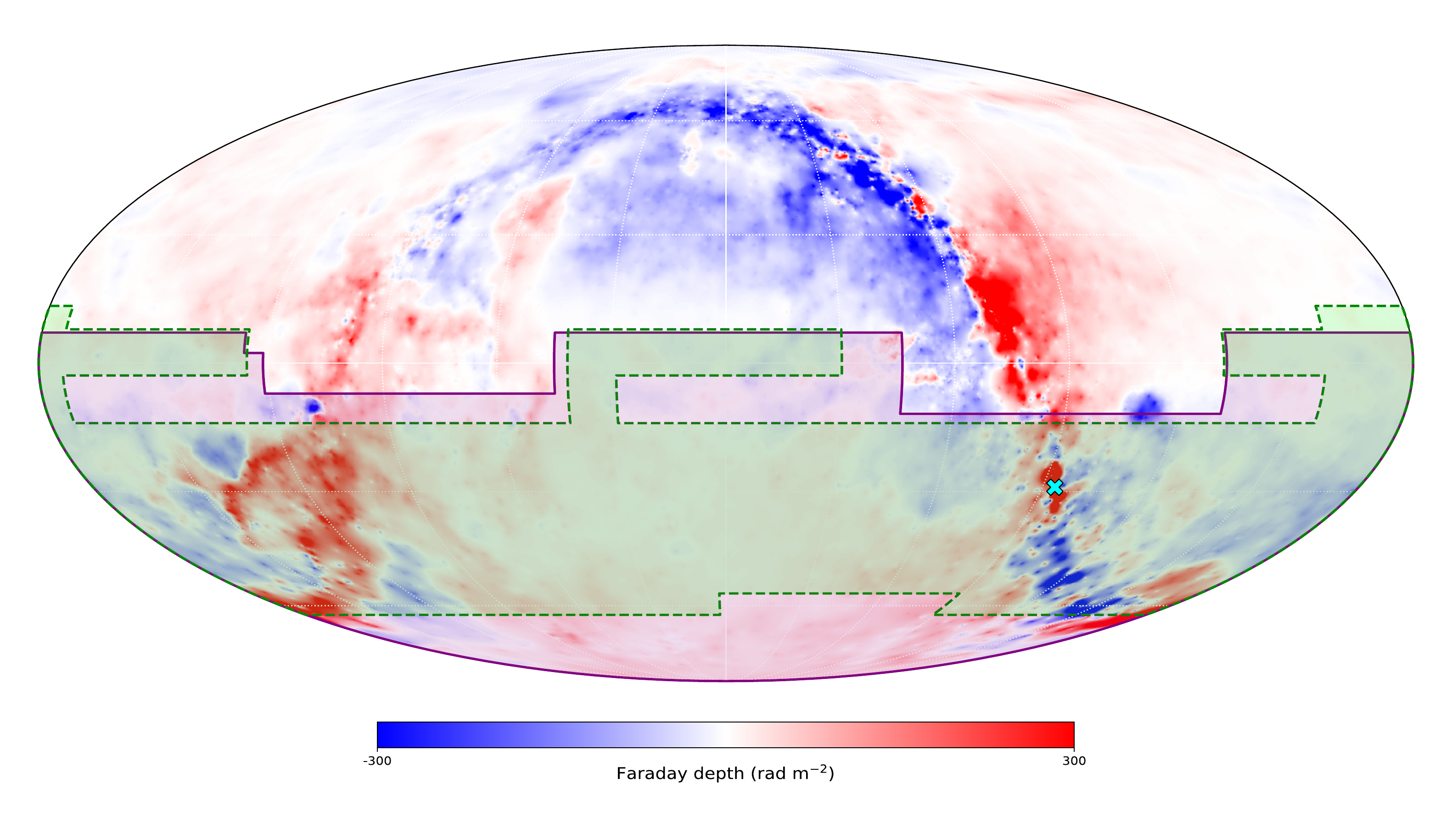}
\caption{Planned sky coverage of POSSUM. The coverage of the band 1 (800--1088\,MHz) component, commensal with the EMU survey, is shaded purple, and the  band 2 (1296--1440\,MHz) component, commensal with the WALLABY survey, is shaded green. (Regions covered by both components appear dark green.) The location of the Galactic Centre is marked by the cyan cross. The background greyscale map in the top panel is the reprocessed \cite{1982A&AS...47....1H} 408\,MHz all-sky synchrotron emission map \citep{2015MNRAS.451.4311R} in linear colour scale, while the background colour map in the bottom panel represents the Milky Way contributions to the Faraday rotation experienced by the linearly polarised emission from extragalactic sources derived by \cite{2022A&A...657A..43H}. Both maps are in equatorial coordinates centred at (J2000) $\alpha = 0^{\rm h}$, $\delta = 0^\circ$ and in Mollweide projection.}
\label{fig:survey_area}
\end{figure*}

\subsection{Survey Design} \label{sec_survey_design}

To achieve the science goals outlined in \S\ref{sec_sci} \emph{and} ensure POSSUM's broad serendipitous discovery potential and legacy value, we identified four key survey characteristics determined by the observational setup (as opposed to the intrinsic imaging capabilities of the array) that  define the capability of POSSUM:

\begin{itemize}
    \item The average sky density of detected linearly polarised background sources;
    \item The RM precision of the detected polarised sources;
    \item The ability to detect and characterise `Faraday complexity', that is, variation of RM within or across the source; and
    \item The total sky coverage, and the overlaps with complementary radio surveys (see \S\ref{sec_others}) and with ancillary multi-wavelength data.
\end{itemize}

These survey characteristics are controlled by the $\lambda^2$ coverage through frequency band choice, target sensitivity through per-field integration time choice, and the selection of sky area covered. 

The choice of wavelength coverage involved significant trade-offs. Longer wavelengths improve RM precision due to the proportional relationship between polarisation angle and $\lambda^2$, and increase source fluxes, following $\lambda^{0.75}$, which can potentially yield more detected sources. However, these benefits come at the cost of reduced angular resolution (which impacts the study of intrinsic emission from resolved objects) and increased differential Faraday rotation both within sources and at scales below the observing resolution (leading to depolarisation; e.g. see \citet{Burn1966, Tribble1991, Sokoloff1998}). This depolarisation becomes particularly pronounced at wavelengths longer than 30 cm (frequencies below 1 GHz), where most bright sources are substantially affected \citep[e.g.,][]{Conway1983}. Balancing these trade-offs required careful consideration of POSSUM's scope and science priorities, alongside empirical measurements of polarised source density and array performance across the available observing bands.

To inform this decision, a pilot survey was conducted, confirming that a greater density of polarised sources is detected in ASKAP's band 1 (0.7--1.2 GHz) compared to band 2 (1.15--1.44 GHz). This difference was particularly pronounced due to the negative impact of RFI from global navigation satellites in the 1.15--1.3 GHz range, which significantly reduced overall sensitivity. The low prevalence of RFI in band 1 also allowed for precise frequency tuning to avoid RFI almost entirely and mitigate system temperature roll-up at the lower end of the band. As a result, the decision was made to conduct the main survey commensally with EMU \citep[][Hopkins et al., 2025, {\it in press}]{2011PASA...28..215N} in ASKAP's band 1, tuned to 800--1088 MHz.

While the choice of frequency tuning was flexible, our imaging strategy faced significant constraints. Modern spectropolarimetric analysis ideally balances finer image cube channelisation, which enhances the recovery of polarised emission with large |RM| values, against coarser channelisation, which reduces data processing and storage demands while improving image deconvolution. However, due to the current limitations of \texttt{ASKAPsoft}, POSSUM's band must be imaged and deconvolved at its native resolution, with each 1 MHz channel processed individually across the frequency range. This approach, though currently necessary, is suboptimal and constrains image quality in certain respects, particularly near bright extended objects. Nevertheless, the archiving of POSSUM visibilities offers the potential to apply alternative custom calibration and imaging pipelines (e.g., \citealp{2023PASA...40...40T}) that could mitigate these issues. Additionally, ongoing developments in \texttt{ASKAPsoft} may eventually address these limitations in future data products, which we plan to highlight in subsequent data releases should this occur.

POSSUM's band 1 observations use the \texttt{closepack36} \citep{2021PASA...38....9H} beam footprint with a beam spacing (pitch) of $0\fdg9$ to balance between the telescope's field of view ($\approx 30\,{\rm deg}^2$), the spatial uniformity of sensitivity, and the wide-field instrumental polarisation response. The planned sky coverage is shown in  Figure~\ref{fig:survey_area}, including a total area of 20,630\,deg$^2$ (50.0\% of the sky) that is presently constrained by the planned survey load for ASKAP over a 5-year period. The entire sky south of declination $\delta \approx -11^\circ 40^{\prime}$ will be covered, with additional sky coverage extending northward to $\delta \approx -7^\circ 0^\prime$ within a J2000 right ascension (RA) range of $03^{\rm h} 05^{\rm m}$ to $08^{\rm h} 05^{\rm m}$, to $\delta \approx +2^\circ 20^{\prime}$ within a RA range of $08^{\rm h} 05^{\rm m}$ to $08^{\rm h} 25^{\rm m}$, and to $\delta \approx +7^\circ 0^{\prime}$ within both RA ranges of $08^{\rm h} 25^{\rm m}$ to $15^{\rm h} 15^{\rm m}$ as well as $20^{\rm h} 55^{\rm m}$ to $03^{\rm h} 05^{\rm m}$ (through $00^{\rm h}$). The irregular declination limit maximises overlap with extragalactic multi-wavelength survey projects.\footnote{In particular, the POSSUM band 1 equatorial coverage from $20^{\rm h} 55^{\rm m}$ to $03^{\rm h} 05^{\rm m}$ through $00^{\rm h}$ is designed to overlap with the Sloan Digital Sky Survey (SDSS) Extended Baryon Oscillation Spectroscopic Survey (eBOSS) emission line galaxy (ELG) cosmology programme \citep{2020ApJS..249....3A}. Meanwhile, the equatorial coverage from $08^{\rm h} 05^{\rm m}$ to $15^{\rm h} 15^{\rm m}$ is purposed to cover the equatorial regions of the Galaxy and Mass Assembly (GAMA) project \citep[in particular, GAMA09, GAMA12, and GAMA15;][]{2022MNRAS.513..439D}. See Hopkins et al., 2025, {\it in press} Figure 4.} The Galactic plane is covered at longitudes $\ell < 19\fdg 8$ and $\ell > 226\fdg 1$. The survey area has been divided into 853 fields following the standard all-sky tiling scheme of ASKAP in equatorial coordinates \citep[see \S9.11 of][]{2021PASA...38....9H}. The integration time per tile is set to 10\,hr, yielding an expected full-band sensitivity of $\sim 18\,\mu{\rm Jy}/{\rm beam}$ at 943\,MHz. For the 692 southern fields at $\delta < -11^\circ 40^\prime$, each tile will be observed in a single 10\,hr session. The remaining 161 northern fields will each be observed in two 5-hr sessions, due to the limited time per day that these fields will be above the ASKAP horizon. The entire survey in band 1 will therefore consist of 1014 observations, or scheduling blocks (SBs).

The band 1 survey will be complemented by ancillary observations in band 2 (1296--1440\,MHz), obtained through a full polarisation processing of the continuum data from the WALLABY survey \citep{2020Ap&SS.365..118K}. The WALLABY survey will cover a sky area of 15,470\,deg$^2$ (37.5\% of the sky) using the \texttt{square\_6x6} PAF footprint with a pitch of $0\fdg9$ at two interleaved positions, totalling 552 fields with 1104 pointings. As outlined by the green-shaded region in Figure~\ref{fig:survey_area}, the band 2 sky coverage will approximately include the entire area within $\delta = -13^\circ 50^\prime$ to $-56^\circ30^\prime$, as well as a southern extension to $\delta = -62^\circ 50^\prime$ within RA range of $00^{\rm h} 10^{\rm m}$--$18^{\rm h} 05^{\rm m}$, a northern extension to $\delta = +7^\circ 45^\prime$ within RA range of $01^{\rm h} 55^{\rm m}$--$02^{\rm h} 45^{\rm m}$, another northern extension to $\delta = +13^\circ 10^\prime$ within RA range of $11^{\rm h} 35^{\rm m}$--$13^{\rm h} 30^{\rm m}$, and two equatorial bands within $\delta = -2^\circ 50^\prime$ to $+7^\circ 45^\prime$ within RA ranges of $08^{\rm h} 20^{\rm m}$ to $15{\rm h} 20^{\rm m}$ as well as $22^{\rm h} 00^{\rm m}$ to $02^{\rm h} 45^{\rm m}$ (through $00^{\rm h}$). The integration time per field (with the two interleaves combined) will be $16\,{\rm hr}$, resulting in a sensitivity of $\sim 19\,\mu{\rm Jy}$ per $10^{\prime\prime}$ beam.

We note that the frequency and sky coverage of bands 1 and 2 are also well-suited for complementary multi-GHz broadband follow-up using the Australia Telescope Compact Array (ATCA), whose lowest frequency is $1.4\,{\rm GHz}$ \citep{2011MNRAS.416..832W} when accounting for RFI, and for complementary polarisation surveys currently being conducted with other instruments, as discussed in \S\ref{sec_other_pol}.

\subsection{Expected Survey Yield}
\label{sec_survey_yield}

The POSSUM survey is designed to produce three key outputs: (1) a dense grid of polarised extragalactic radio sources, (2) intrinsic polarisation measurements of individual resolved Galactic and extragalactic objects, and (3) mapping of Galactic diffuse synchrotron emission. The scientific objectives for resolved and diffuse polarisation products are detailed in Sections \ref{sec_sci} and \ref{sec_single_dish_surveys}, respectively. In this section we focus on the expected characteristics of POSSUM's RM grid.

Previous polarisation studies suggested that POSSUM's band 1 survey would yield 30--50 RMs~deg$^{-2}$ (e.g., \citealp{RO2014erratum,Loi2019}) based on a conservative $8\sigma$ detection threshold \citep{Macquart2012} of 144~$\mu$Jy/beam. This projection is now supported by results from POSSUM's pilot survey \citep{2024AJ....167..226V}, which achieved an RM density of 42~deg$^{-2}$ in an extragalactic field and 20~deg$^{-2}$ in a Galactic-plane field, with densities slightly decreasing to 35 and 14 RMs~deg$^{-2}$, respectively, for Faraday-simple sources (i.e., those showing no evidence of multiple polarised emission components; see e.g., \citealp{Anderson2015}). Consequently, across the survey region, we expect to measure $(6.2-10.3) \times 10^5$ RMs for robustly detected polarised radio sources. We emphasise, however, that POSSUM will catalogue polarisation data for \emph{all} sources catalogued by the EMU survey in Stokes $I$, irrespective of their measured polarised signal-to-noise ratio (see \S\ref{sec:obsproducts}), allowing users to set their own detection thresholds and balance the trade-off between RM grid density and reliability.

With band 1's frequency coverage, the full width at half maximum (FWHM) of the rotation measure spread function (RMSF) in RM synthesis is $58~{\rm rad\,m}^{-2}$. An $8\sigma$ polarized signal-to-noise cutoff allows for an RM precision of approximately 3~rad~m$^{-2}$ at the detection limit, with mean and median RM uncertainties of 2.1~rad~m$^{-2}$ and 1.6~rad~m$^{-2}$, respectively, across the detected source population \citep{2024AJ....167..226V}. These uncertainties can be compared to the intrinsic RM scatter of extragalactic point sources at this resolution and frequency, which is approximately 5~rad~m$^{-2}$ \citep[e.g.,][]{Schnitzeler2010,2024MNRAS.528.2511T}.

For sky areas with both band-1 and band-2 coverage, the combined sensitivity of $\sim 12\,\mu{\rm Jy/beam}$ is expected to yield a polarised source density of $45$--$60\,{\rm deg}^{-2}$ at $8\sigma$ threshold, with a measured value of 48 ${\rm deg}^{-2}$ (37\,deg$^{-2}$ for Faraday-simple sources) delivered by the Pilot Survey \citep{2024AJ....167..226V}. The RMSF FWHM of $36\,{\rm rad\,m}^{-2}$ will yield a median RM precision of $\approx 1\,{\rm rad\,m}^{-2}$ across all detected polarised sources.

\begin{figure}[ht!]
    \centering
    \includegraphics[width=\textwidth]{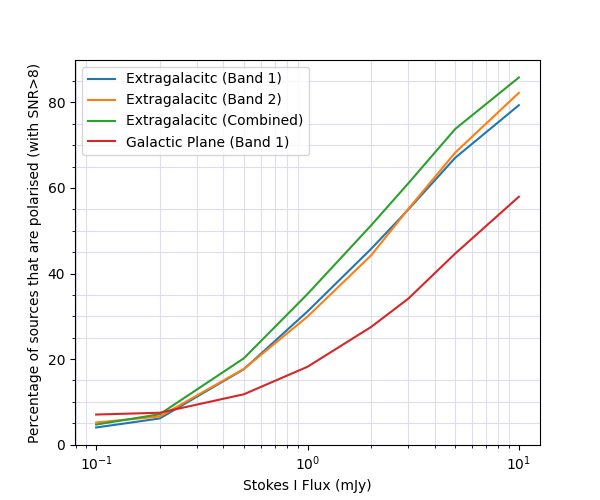}
    \includegraphics[width=\textwidth]{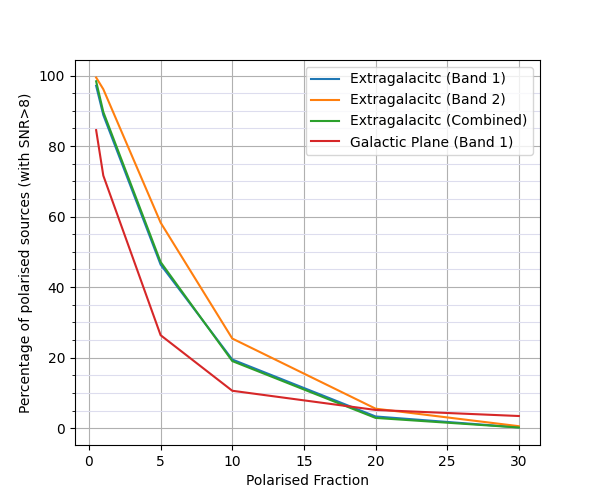}
    \caption{The expected percentage of polarised sources for Stokes I fluxes above various thresholds (top) and percentage of polarised sources for various polarised fractions (bottom) computed using the \cite{2024AJ....167..226V} prototype POSSUM catalogues.}
        \label{fig:polstats}
\end{figure}

From the \cite{2024AJ....167..226V} study, we anticipate approximately 16\% of band-1 polarised components in extragalactic regions will be complex. Near the Galactic plane, this number can be as high as 33\%. Using the catalogues published in that same study, we predict that around 4-5\% of all EMU Stokes I components will be polarised with a signal-to-noise ratio $>8$. For Stokes I components brighter than 10~mJy, these catalogues show that up to 85\% of sources are polarised. Of all the polarised sources, around 90\% have polarised fraction $>$1\% and around 50\% have polarised fractions $>$5\%. These values are summarised in Fig.~\ref{fig:polstats}. The \cite{2024AJ....167..226V} study also compares the POSSUM RMs to existing values in the literature. Comparisons are challenging and limited due to the low source density of the existing catalogues, but they find generally good agreement where comparisons can be made.

\subsection{Survey Timeline}

POSSUM began trial survey observations in November 2022, and commenced the main survey in May 2023. The survey is expected to run for five years, concluding around May 2028. To indicate the typical pace of observations and data availability, by 16 July 2024, 200 of the 1014 planned band 1 scheduling blocks (SBs) had been validated and released, with an additional 19 observed and awaiting processing. For band 2, 24 out of 1104 SBs had been validated, with 2 more observed and pending processing. The survey status is regularly updated online.\footnote{\url{https://www.mso.anu.edu.au/~cvaneck/possum/}}

\subsection{Performance Relative to Other Polarisation Surveys} \label{sec:compare}

\begin{figure}[ht!]
    \centering
    \includegraphics[width=\textwidth]{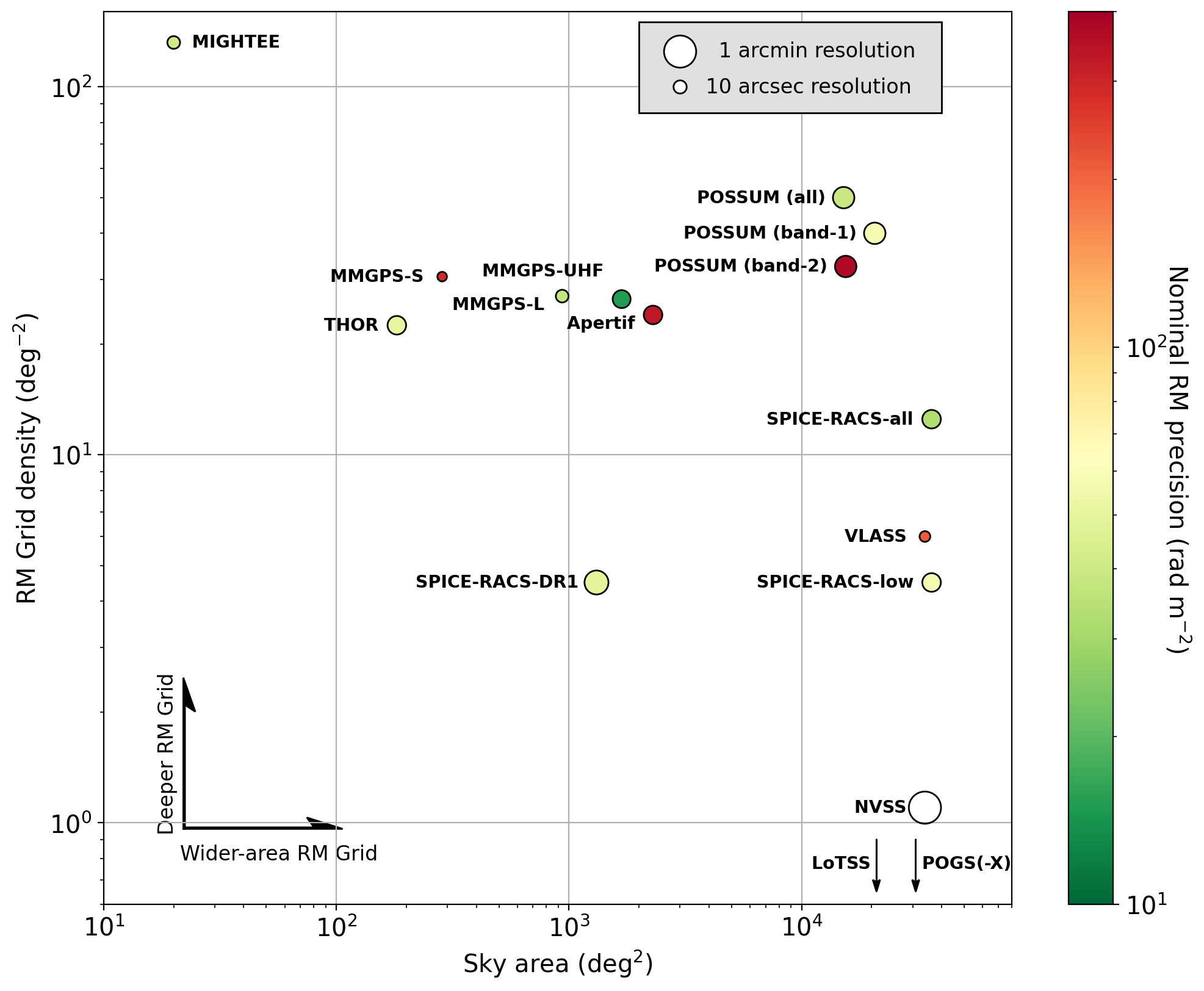}
    \caption{Comparison of POSSUM's performance with other surveys listed in Table~\ref{tab:polsurveys} across several key metrics. The RM precision is depicted by the colour of each circle, according to the logarithmic scale on the right. The size of each circle corresponds linearly to the angular resolution of the respective survey. While NVSS is included for reference, its RM precision is not displayed, as it is more suitable for characterizing broadband surveys. Low-frequency surveys like LoTSS, POGS, and POGS-X lie off the bottom of the plot due to their considerably lower RM Grid density compared to mid-frequency surveys, despite offering much better RM precision.}
        \label{fig:turfplot}
\end{figure}

We now compare POSSUM's design (\S\ref{sec_survey_design}) and  performance (\S\ref{sec_survey_yield}) to other past and present radio polarisation surveys, to underscore POSSUM's unique potential for advancing studies of cosmic magnetic fields. Key parameters for these surveys are listed in Table~\ref{tab:polsurveys} and summarised in Figure~\ref{fig:turfplot}. 

As highlighted in \S\ref{sec_sci}, the polarisation re-analysis of the NVSS by \citet{2009ApJ...702.1230T} ushered in a revolution in RM grid science. However, the NVSS RM grid is relatively sparse by modern standards, with a density of approximately 1 RM deg$^{-2}$. Additionally, its median RM uncertainty of around 10 rad m$^{-2}$ is comparable to the signal strengths of the very targets we now aim to study (see \S\ref{sec_sci}). Finally, it suffers from issues of precision and reliability due to the limited number of spectral channels used (e.g., \citealp{2019MNRAS.487.3432M,2019MNRAS.487.3454M}).

Subsequent surveys have improved on the NVSS in key respects, often dramatically so, but these advancements typically involve trade-offs in other critical areas, such as coverage area, resolution, or sensitivity. For example, \cite{2014ApJ...785...45R} conducted a 1.5-GHz polarisation survey with a detection threshold of 14.5~$\mu$Jy~beam$^{-1}$, yielding a grid density of approximately 100~RMs~deg$^{-2}$ at $10''$ resolution, but only covering an area of 0.27 square degrees (0.00065\% of the sky).\footnote{\citet{2014ApJ...785...45R} found a significantly lower number density at their full 1\farcs 6 resolution, suggesting that polarised sources are often extended over several arcseconds.} The MeerKAT International GHz Tiered Extragalactic Exploration (MIGHTEE) survey \citep{2024MNRAS.528.2511T} aims to provide polarisation data with a sensitivity of a few $\mu$Jy/beam and a resolution of $18''$, though its coverage is limited to 20~deg$^2$ (0.048\% of the sky). The Very Large Array Sky Survey (VLASS) \citep[VLASS;][]{Lacy2020} has covered the entire sky north of declination $-40^\circ$ (33,827 deg$^2$, or $\sim$82\% of sky) at 2--4 GHz with 5$''$ FWHM in polarisation, significantly improving upon NVSS in resolution and enabling the mapping of RM structure across radio galaxies, albeit with substantially greater RM uncertainties than NVSS, and a largest angular scale of recoverable emission dramatically lower than most other large-area $\sim$GHz-frequency surveys. At lower frequencies, surveys conducted by the Low-Frequency Array (LOFAR) and Murchison Widefield Array (MWA) together cover the entire sky at meter wavelengths (see \S\ref{sec_low_freq}), achieving high RM precision, though depolarisation effects result in a much lower RM grid density compared to higher frequency surveys.

POSSUM will vastly outperform the NVSS RM catalogue and polarimetric imaging in several key areas, including RM sky density and RM uncertainties, improving them by factors of approximately 35 and 10, respectively. This will provide an RM grid containing hundreds of times more information. Additionally, POSSUM offers enhanced spatial resolution at $20''$, and its fully-sampled band will enable the detailed identification and characterization of Faraday complexity---i.e., multiple polarised emission components with distinct RMs or components that span a range of RMs. The exceptional image quality, surface brightness sensitivity, the approximately 30 arcminute largest angular scale of recoverable emission provided by ASKAP, and the excellent wide-field polarisation purity achieved in routine end-to-end data processing (see \S\ref{sec_ASKAP}) provides peerless new capabilities for mapping intrinsic polarisation in large samples of resolved objects.

The unique combination of deep, large-area observations processed in an automated, end-to-end manner, covering the previously under-sampled southern sky---now a focal point for some of the world's most advanced telescopes across all wavelengths---will make POSSUM's survey data an invaluable legacy resource. This resource will remain unsurpassed until science-ready data products from all-sky SKA surveys become available.

\begin{table*}
	\centering
	\caption{POSSUM compared to other major modern interferometric polarisation surveys.}
	\label{tab:polsurveys}
	
\setlength{\tabcolsep}{4pt} 

\begin{tabular}{lc>{\centering}p{2.2cm}D{.}{.}{-1}rcrrrcrcc}
		\hline
		\hline
		Survey & Frequency $^a$ & Coverage & \text{\% of sky} & 1-$\sigma$ noise $^b$ & LAS$^c$ & Beam $^d$  & RM density & $\delta\phi$ $^e$ & $\phi_{\rm max}$ $^e$ & $\phi_{\rm max-scale}$ $^e$  & Ref \\ 
		&MHz&& \% & $\mu$Jy beam$^{-1}$ & arcmin & arcsec & deg$^{-2}$ & rad m$^{-2}$& rad m$^{-2}$&rad m$^{-2}$&\\
		\hline
        POSSUM (band-1)  & 800--1088  & Fig.~\ref{fig:survey_area} & 50 & 18 & 30 & 20 &30--50& 58& 8100 & 41 & [1]\\
        POSSUM (band-2)  & 1296--1440  & Fig.~\ref{fig:survey_area} & 37.5 & 19 & 21 & 20 &25--40& 370 &25000& 73 & [1]\\
        POSSUM (all)  & 800--1440$\dagger$  & Fig.~\ref{fig:survey_area} & 36.75 & 18 & 26 & 20 &40--60& 39 &25000& 73 & [1]\\
        Apertif $^f$ & 1292--1442 & Irregular & 5.57 & 16 & 15 & 15 & 24 & 360 & 32000 & 73 & [2]\\
        LoTSS $^g$   & 120--168   & $-1^\circ <\delta< +90^\circ$ & 50.75 & 70 & 43 & 20 &0.43&1.2&290&1.0 & [3]\\
        MIGHTEE & 880--1680$\dagger$  & Small fields  & 0.048 & 2 & 19 & 18 &99--165 & 45 & 8100 & 99 & [4]\\
        MMGPS-UHF&544--1088$\dagger$& $-62^\circ < \ell < +15^\circ$, \newline  $|~b~| < 11^\circ$& 4.11 &$\sim$40 & 38 & 14 & 20--33 & 17 & 39000 & 41 & [5]\\
        MMGPS-L&856--1712$\dagger$ & $-100^\circ < \ell < -10^\circ$, \newline
        $|~b~| < 5.2^\circ$
        & 2.27 & 25 & 24 & 9 & 22--36 & 41 & 98000 & 100 & [5]\\
        MMGPS-S&1968--2844$\dagger$ & $-80^\circ < \ell < +15^\circ$,\newline
         $|~b~| < 1.5^\circ$ & 0.69 & 15 & 13 & 5 & 23--38 & 310 & 630000 & 280 & [5]\\
        NVSS  & 1365--1435 & $-40\degr < \delta < +90\degr$ & 82.25 & 450 & 13 & 45 & 1.1 & -- $^h$ & 640 & -- & [11]\\
        POGS    & 169--231 & $-82^\circ<\delta<+30^\circ$ & 74.5 & 1500 & 62 & 300 & 0.02 & 2.6 & 1900 & 1.9 & [6]\\
        POGS-X    & 169--231 & $-90^\circ<\delta<+30^\circ$ & 75 & 1200 & 62  & 120 & 0.02 & 2.6 & 1900 & 1.9 & [7]\\
        SPICE-RACS DR1 & 744--1032 & Spica Nebula & 3.18 & 150 & 32 & 25 &4.5& 49 & 6700 & 37 & [8] \\
        SPICE-RACS-low $^i$ & 800--1088 & $-90\degr < \delta < +49\degr$ & 87.75 & 170 & 30 & 15 &7.3& 58 & 980 & 41 & [8] \\
        SPICE-RACS-all $^i$ & 800--1800$\dagger$ & $-90\degr < \delta < +49\degr$ & 87.75 & $\sim$100 & 26 & 15 &9--16& 34 & 4500 & 113 & [8] \\
        THOR/THOR-GC    & 1000--2000$\dagger$ & $-6^\circ < \ell < +67^\circ$, \newline
         $|~b~|<1.25^{\circ}$& 0.45 & 35 & 12 &15&3.4& 56 & 16000 &140&[9]\\
        VLASS $^j$   & 2000--4000$\dagger$ & $-40^\circ<\delta<+90^\circ$& 82.25 & 69 & 1 & 5&6& 220 & 2000 & 560& [10] \\
		\hline
	\end{tabular}

\begin{tablenotes}
\item[a] Lowest and highest usable frequencies. Bands containing substantial gaps (e.g., due to RFI) are marked with $\dagger$.
\item[b] Noise in Stokes $Q$ and $U$.
\item[c] Largest Angular Scale of recoverable emission, estimated as LAS $= 0.6\lambda_{\text{c}}/B_{\text{min}}$, where $\lambda_{\text{c}}$ is the wavelength at the centre of the band, and $B_{\text{min}}$ is the minimum baseline length.
\item[d] FWHM of the sky beam used for polarisation analysis; $\ge$ resolution of lowest frequency channel, so usually lower than associated Stokes $I$ images.
\item[e] $\delta\phi = 3.8/\Delta\lambda^2$: FHWM of RM spread function; $\phi_{\rm max} = \sqrt{3}/\delta\lambda^2$: maximum measurable RM; $\phi_{\rm max-scale} = \pi/\lambda_{\rm min}^2$: maximum detectable scale of extended structure in Faraday space.
\item[f] Apertif re-tuned to 1220-1520 MHz (1292-1520 MHz usable) for the final year, to reduce RFI contamination.
\item[g] A later LoTSS data release will have $6''$ angular resolution.
\item[h] NVSS did not employ Faraday synthesis as only two frequency channels were available. For comparison, the median value of the tabulated NVSS RM uncertainties is ten times larger than for the POSSUM band 1 survey.
\item[i] SPICE-RACS spatial resolution is roughly declination-dependent.
\item[j] VLASS 
$\phi_{\rm max}$ is for the initial data release;
final 16-MHz channels will give $\sim$16,000~rad m$^{-2}$.
\medskip

 References: [1] This paper; [2] \citet{adebahr22}, \citet{2022A&A...667A..38A}; [3] \citet{shimwell22,2023MNRAS.519.5723O}; [4] \citet{2024MNRAS.528.2511T}; [5]~\citet{2023MNRAS.524.1291P}; [11] \citet{2009ApJ...702.1230T}; [6] \citet{2018PASA...35...43R,2020PASA...37...29R}; [7] Zhang et al. (in prep); [8] \citet{2023PASA...40...40T};
 [9] \citet{beuther16}, \citet{2022ApJ...939...92S}; [10]
\citet{Lacy2020}.

\end{tablenotes}
\end{table*}

\section{DATA PROCESSING AND SOFTWARE PIPELINES}
\label{sec_processing}

POSSUM will operate several data processing pipelines that will take the data produced by ASKAP and will generate science-ready data products suitable for some kinds of polarisation analysis, as well as final data products such as polarisation catalogues. A full detailed description of the POSSUM pipelines is left to a follow-up paper (Van Eck et al., in prep) to be published at or before the first data release, but below we summarize both the major steps in the ASKAP pipelines that pertain to polarisation processing and the subsequent POSSUM-specific pipelines that we are developing.

\subsection{Observatory Pipelines}
\label{sec_imaging}
The ASKAP observatory maintains and operates data processing pipelines to produce the basic survey data products. The observatory's POSSUM-specific pipeline handles all aspects of interferometric calibration and imaging, producing images, data cubes, and related products. This pipeline is intended to operate in a near-real-time mode, with calibration and imaging being performed as quickly as possible after an observation is complete. As a result, each observation is processed through this pipeline fully independently, even where multiple observations cover the same sky area (i.e., interleaved observations).

The bandpass is derived from observations of the unpolarised calibrator source PKS~B1934--638. For observations with visibilities recorded at higher spectral resolution (i.e., in observations commensal with WALLABY), the calibrated visibilities are averaged to 1-MHz coarse resolution channels after bandpass calibration. Flagging is applied to the bandpass observations prior to computing bandpass solutions, and to the target data after applying bandpass solutions.

Post primary calibration, the on-axis leakages for ASKAP are usually less than 1\% (see \S\ref{sec_ASKAP}). Data on the bandpass calibrator source are used to derive the on-axis leakage spectra per antenna per formed ASKAP beam, which are applied to the target science data after the time-dependent gains are refined using self calibration. This calibration corrects the on-axis instrumental leakages in ASKAP data to better than 0.1\%.

Intermittent observation of secondary calibrators to refine time-dependent gain variations is not feasible for multiple formed beams. For ASKAP, gain variations for each beam are derived using the iterative technique of self-calibration. For POSSUM data, a single amplitude and phase self-calibration iteration is used. 
This process leads to a small but non-zero astrometric shift with respect to reference catalogues, an issue which is described in detail in the EMU overview paper (Hopkins et al., 2025, {\it in press}).

Images are generated and {\em CLEAN}ed independently per beam. Stokes $IQUV$ spectral cubes are produced, with each channel and Stokes parameter imaged and {\em CLEAN}ed independently. Meanwhile. Stokes $I$ and $V$ multi-frequency synthesis (MFS) images, using the full observation bandwidth, are made for EMU and WALLABY (see \S\ref{sec:other_ASKAP}).

The synthesised beams or point spread functions (PSFs) corresponding to the different formed ASKAP beams are not necessarily identical, because each beam samples slightly different spatial frequencies (due to having different target directions) and because there is different flagging for each beam. The smallest common PSF size for all 36 formed beams in a given observation is calculated on a per-channel basis and all beams are convolved to have this PSF. Spectral variations in the PSF size are preserved, so higher frequency channels will typically have a smaller PSF.

The images for each formed beam are linearly mosaicked together to form a single set of image and cube products, each covering the full area of a single observation. The mosaicking uses full-Stokes models of the primary beam derived from holographic observations of the point source PKS~B0407--638 to apply corrections for both the sensitivity pattern of the primary beam and the off-axis leakage of Stokes $I$ into $Q$, $U$, and $V$.

After mosaicking, the source finding software {\em Selavy} \citep{Whiting2012} is run on the Stokes $I$ MFS image, producing catalogues of total intensity source components and emission islands.\footnote{Components are individual sources modelled as the telescope's synthesised beam shape, and islands are groups of nearby components likely associated with the same physical structure.} These Stokes $I$ catalogues will report fitted and deconvolved sizes along with fitted peak and total intensities. The 1D POSSUM pipeline will be run on each Stokes $I$ component and thus the polarised and total intensity values can be linked between the catalogues. Several additional tools, such as those that generate the data quality reports used for validation, are also run on the mosaicked images and data cubes. The mosaicked images, cubes, and other ancillary data are then deposited in the ASKAP data archive.

\subsection{Archiving, Quality Control, and Validation}\label{sec_QA}

The observatory pipelines deposit data into the CSIRO ASKAP Science Data Archive (CASDA; \citealp{Huynh2020}) and assign an ``Unreleased'' state that makes them visible only to members of the relevant survey teams.

Once the data have been uploaded to CASDA, the survey team makes an initial assessment to evaluate whether the data are of sufficient quality to enable science; if so, the data are released according to the ASKAP data release policy.\footnote{\url{https://www.atnf.csiro.au/projects/askap/ASKAP\_Publication\_Policy.pdf}}

Since the output data cubes are very large (up to 180~GB),  manual download and inspection of each observation is difficult. The quality assessment is therefore performed using an HTML-format report that is created within the ASKAP pipeline. This report extracts a variety of numerical statistics, diagnostic plots, polarisation spectra of bright sources, JPEG versions of the images, and GIF movies that cycle through the cube channels, all of which are used to perform an initial manual inspection of the data.

\subsection{Science-Ready Processing Pipeline}\label{sec_AusSRC}
The observatory pipeline model of near-real-time processing of individual observations has a few limitations that prevent it from performing all the processing needed to produce the final survey data products: it cannot perform processing that requires multiple observations or that requires additional data that are not yet available immediately after an observation is completed. POSSUM has developed an additional science-ready processing pipeline that performs steps that cannot be accomplished with the observatory pipelines, to get the data to a final science-ready state (i.e., requiring no additional processing steps prior to beginning scientific analysis).

The science-ready processing pipeline has been built and operated as a partnership between POSSUM team and the Australian SKA Regional Centre (AusSRC), as part of the AusSRC's interest in developing expertise with SKA-like large-data pipelines. The pipeline performs four primary steps: convolution to survey resolution, correction for ionospheric Faraday rotation, tiling and reprojection, and mosaicking of overlapping/adjacent observations.

Convolution to survey resolution is required for three reasons: to enforce uniformity across the survey area, which simplifies some types of analysis that can be affected by resolution; to allow adjacent observations to have a common PSF to allow them to be mosaicked together; and to ensure that all frequency channels have a common PSF prior to spectral analysis. The convolution to survey resolution could be performed in the observatory pipeline, but it was decided that there is benefit to having an archived version of each observation at the best possible resolution, since this would allow users to use the better resolution data (at the cost of additional processing by the user) if needed for their specific scientific needs.

The survey resolution was chosen to be $20''$, as this was found to be a reasonable compromise between preserving the best resolution possible (approximately $18''$, based on the typical PSF size at the lowest frequency of 800 MHz) and keeping channels or observations for which data flagging has resulted in slightly worse PSFs. In individual observations, any channels so affected by flagged visibilities that the resolution is worse than $20''$ are flagged out in the data cubes; if more than 10\% of channels are affected, an observation will be rejected during validation and will require re-observation.

Ionospheric correction is performed to subtract the Faraday rotation caused by the Earth's ionosphere, which would otherwise contaminate the astrophysical Faraday rotation and introduce systematic errors. This correction would ideally be applied to the visibilities prior to imaging using external measurements of ionospheric electron content, which are published on the internet with latency of hours to days. However, the ASKAP observatory's goal of near-real-time processing does not allow ionospheric correction prior to imaging, preventing a full correction of the time-dependent ionospheric Faraday rotation. Imaging without correcting the time-dependent effects introduces some amount of depolarisation, but at POSSUM frequencies this effect is typically much smaller than a 1\% loss of polarised intensity, so it is primarily the systematic shift in RMs and polarisation angles that requires correction. We have developed a software tool\footnote{\url{https://frion.readthedocs.io/en/latest/}} to calculate and apply a time-averaged correction to the Stokes $Q$ and $U$ cubes. Typical corrections are of order $1~\mathrm{rad\,m^{-2}}$.

Tiling the data (i.e., breaking it into multiple files each covering smaller sky area) is necessary, as the large (>150 GB) FITS cubes produced by the observatory are difficult to work with outside of a high-performance computing facility. We have developed a tiling scheme based on the HEALPix \citep{Gorski2005} pixel grid: each tile is defined as the sky area covered by a single pixel of the HEALPix grid with $N_{\rm side} = 32$; each tile thus has a sky area of approximately 3.36~deg$^2$ (or approximately $110'$ per side). Each tile contains $2048\times 2048$ pixels with an approximate linear size of 3\farcs 22 (equivalent in size and position to HEALPix pixels of $N_{\rm side}$ = 65 536) in the HPX projection.\footnote{It is important to note that the tiles are in the HPX projection \citep{2007MNRAS.381..865C} and defined using the borders of HEALPix pixels, but the resulting files are {\em not} stored as HEALPix files (i.e., as 1D tables of pixel values). The files are standard FITS images (2D or 3D arrays) and can be interacted with using most conventional FITS image viewers.} 
The principle is similar to the HiPS format \citep{Fernique2017}, but without the hierarchical scheme of lower resolution versions present in HiPS. Our tile size is approximately 4.2 GB, small enough to be easily manipulated even on reasonably modern laptops. The survey area spans 6319 tiles in band 1 and 4985 tiles in band 2.

Finally, tiles from overlapping or adjacent observations are linearly mosaicked together, which mitigates the noisy edges around each observation and provides a more uniform sensitivity across the boundaries of observations. These final tiles are downloaded and ingested to the POSSUM collection at the Canadian Astronomical Data Centre archive.\footnote{\url{https://www.cadc-ccda.hia-iha.nrc-cnrc.gc.ca/en/search/?collection=POSSUM\&noexec=true}}

The POSSUM analysis pipeline comprises two parallel components. The first focuses on polarisation catalogues and polarisation spectra of source components. This is called the 1D pipeline, since it reduces each source component to a single dimension (frequency). The second component focuses on resolved and extended sources and on diffuse polarised emission. This is the 3D pipeline, using three dimensions: right ascension, declination, and frequency or RM.\footnote{Or Faraday depth. RM and Faraday depth are distinct but related concepts that we do not elaborate on here. For detailed explanations, see \citet{Burn1966} and \citet{2005A&A...441.1217B}.}

\subsection{1D Polarimetry Pipeline}\label{1d_pipeline}

The 1D component of POSSUM's science-ready processing pipeline was built around the idea of providing polarisation information for every single Stokes $I$ source component present in the EMU catalogues, including non-detections. Rather than performing an untargeted search for polarised sources, each EMU source component is assigned a counterpart in the POSSUM catalogue with the same name/ID, allowing the catalogues to be easily joined and giving users direct access to the full power of both surveys together.

For each EMU source component, the Stokes $IQU$\footnote{Stokes $V$ data cubes are generated from POSSUM observations and made publicly available, they are not included in subsequent POSSUM processing or catalogues, as their analysis falls outside the survey's primary scientific scope.} spectra (with corresponding measurement uncertainties) are extracted from the tiles produced as described previously. An estimate of the diffuse foreground/background signal present around each source is also derived, and is subtracted from the on-source spectra; this estimate is stored along with the on-source spectra.

Using these spectra, RM synthesis is performed using the RM-Tools package \citep{2020ascl.soft05003P}\footnote{\url{https://github.com/CIRADA-Tools/RM-Tools}} to transform the observed complex fractional polarised signal as a function of wavelength-squared, $\mathbf{P}(\lambda^2)$, into the ``Faraday dispersion function'' \citep[FDF;][]{Burn1966,2005A&A...441.1217B}. The FDF encodes properties such as the polarised fraction and RM of the strongest polarised emission peak, and can be used to estimate the degree of Faraday complexity.

The polarisation information is then merged with that from the Stokes $I$ source-finding, along with relevant information from the data quality assessment/validation and observation metadata, to create the POSSUM catalogue. At this stage, various flags for properties such as significant polarised intensity or suitability for RM grid analyses are assigned; these will be discussed in detail in the forthcoming data release and pipeline description papers.

\subsection{3D Polarimetry Pipeline}\label{3d_pipeline}

The 3D component of POSSUM's science-ready processing pipeline focuses on the production of products to enable the analysis of spatially-resolved polarised emission, which can include resolved sources as well as diffuse Galactic emission. This pipeline is very straightforward: it performs RM synthesis on every pixel location in the Stokes $IQU$ datacubes, producing FDF cubes matching the area of each frequency cube tile. An FDF characterisation routine is also run producing a series of 2D maps of polarisation properties as a function of sky position, such as peak Faraday depth, polarized intensity and fraction, and more, along with uncertainty estimates on these quantities.

\subsection{Future Processing Possibilities}

Several additional pipelines and products are currently being considered, and will be implemented depending on the level of team/community interest and on available development and computational resources. 
For example, a project is being developed to provide storage and computing resources from the China SKA Regional Centre Prototype (CSRC-P) to process calibrated and leakage-corrected measurement sets to produce image cubes optimised for diffuse polarised emission. Single-dish data from the PEGASUS project (see \S\ref{sec_single_dish_surveys}), when available, can then be merged with these cubes to produce images sensitive to all spatial scales, which will be crucial to study diffuse emission.
Pipelines to perform alternative or additional polarisation analysis of individual sources, such as searches for multiple significant peaks in FDFs, $QU$-fitting to sources with detected polarisation using selected models, or characterisation of partially resolved sources, plus pipelines to combine band-1 and band-2 data, have been discussed and may be developed in the future.

\section{DATA PRODUCTS}
\label{sec_products}

The POSSUM data products can be divided into three groups: observatory products, science-ready survey products, and analysis pipeline products. 

The observatory products are all of the outputs of the ASKAP pipeline, as described in \S\ref{sec_imaging} and \S\ref{sec_QA}. These products are archived on CASDA,\footnote{\url{https://data.csiro.au/domain/casdaObservation}} and data that pass validation by a POSSUM team member immediately become publicly available (see Section \ref{sec_QA}).

The science-ready and analysis pipeline products are produced by the POSSUM pipelines (see \S\S\ref{sec_AusSRC}, \ref{1d_pipeline}, \ref{3d_pipeline}). These products are released to the POSSUM membership as they are produced, and are proprietary to the collaboration for a period of 12 months. By default, these products automatically become public 12 months after archiving, but the release dates may be adjusted to accommodate staged data releases. These products are hosted by the Canadian Astronomical Data Centre (CADC),\footnote{\url{https://www.cadc-ccda.hia-iha.nrc-cnrc.gc.ca/en/search/?collection=POSSUM\&noexec=true}} which provides tools for exploring, visualizing, and downloading these products.

The following subsections describe each set of data products in more detail.

\subsection{Observatory Products} \label{sec:obsproducts}
The primary POSSUM products from the observatory pipeline are continuum channelised cubes for all Stokes parameters, $I$, $Q$, $U$, and $V$, at 1-MHz channel resolution. These are released as large FITS files, covering the full area of a single observation (approximately 6$^\circ$ by 6$^\circ$).The channels are not convolved to a common beam size, but are instead kept at the best resolution possible during the beam mosaicking process (on a per channel basis); the PSF for each channel is stored as a FITS extension in the cube files. The resulting data cubes are typically 192~GB per Stokes parameter for band-1 observations and 81~GB for band 2.

Each observation also has corresponding archival products from EMU or WALLABY. For both surveys, this includes Stokes $I$ and $V$ MFS images, Stokes $I$ component and island catalogues, visibilities (averaged to 1-MHz channels), and validation reports; WALLABY observations also include a Stokes $I$ spectral line cube and additional visibilities.

Despite ASKAP's excellent polarimetric performance as highlighted in \S\ref{sec_ASKAP}, we must raise a caveat: SBs validated and accepted prior to October 5, 2023, had the on-axis leakage correction applied twice in error. This resulted in an expected leakage of Stokes $I$ into $Q$ and $U$ of approximately 1\%. This issue affects 98 SBs, or about 10\% of the total survey area. We will endeavour to reprocess and correct these affected SBs as soon as possible. 

\subsection{Enhanced Data Products}
The pipeline described in \S\ref{sec_AusSRC} produces Stokes-frequency data cubes for Stokes $I$, $Q$, and $U$ (Stokes $V$ is not processed), as well as the EMU Stokes $I$ MFS images convolved to POSSUM resolution, to match the cubes. These files follow the POSSUM HPX-projection tiling scheme (see \S\ref{sec_AusSRC}), resulting in cubes of approximately 4.5~GB for band 1 and 2.3~GB for band 2.

We intend to produce a tool to enable easy creation and download of image/cube cutouts (including stitching/mosaicking across adjacent tiles). However, this has not been developed at the time of writing, and is anticipated for the first data release.

\subsection{1D Pipeline Products}
\label{sec:POSSUM_catalogue}
The primary product of the POSSUM 1D pipeline described in \S\ref{1d_pipeline} is a catalogue of the polarisation properties for every source component that appears in the EMU component catalogue. This catalogue is constructed as a one-to-one match to the EMU catalogue and contains the EMU component names, which allows trivial matching of source components between these two catalogues and give users easy access to the full information provided by both catalogues. The POSSUM catalogue contains information on the basic polarisation properties (e.g., fractional polarisation, RM, polarisation angle), sourcefinder properties (e.g., deconvolved size, peak intensity, integrated flux density), observation metadata (e.g., observation IDs and dates, data quality metrics), and source assessment flags (e.g., polarisation detection/non-detection, Faraday simplicity/complexity, suitability for use in an RM grid). Full descriptions of the pipeline and the resulting catalogue will be presented in the first data release and pipeline overview papers.

The specifics of the archiving and public access to the catalogue are not finalised at the time of writing, but are expected to include a live service allowing queries through a Table Access Protocol (TAP) interface.\footnote{\url{https://www.ivoa.net/documents/TAP/}}

In addition to the catalogue, the 1D pipeline also outputs the Stokes $I$, $Q$, and $U$ spectra extracted for each source component, and the resulting FDFs computed from those spectra. The spectra may be useful for re-analysis with different techniques, or for combining with observations at other frequencies. The spectra are archived per-tile in the PolSpectra\footnote{\url{https://github.com/CIRADA-Tools/PolSpectra}} FITS binary table format \citep{VanEck2023} and will be made available through a TAP-accessible database; the FDFs will similarly be stored in both FITS binary tables and a TAP interface.

\subsection{3D Pipeline Products}
The primary outputs of the 3D pipeline (see \S\ref{3d_pipeline}) are FDF cubes, containing the FDF for each pixel of sky area. These cubes share the same tiling scheme and pixel grid as the Stokes-frequency cubes from which they are derived. The FDF-characterisation part of the 3D pipeline produces a series of maps of various polarisation properties, such as the polarised intensity, peak rotation measure, and polarisation angle of the strongest peak in the FDF, along with maps of the (statistical) measurement uncertainties in these quantities. The complex RMSF is also stored; all of the required information needed to deconvolve the FSF using the RMSF \cite[e.g., using RMCLEAN;][]{2009A&A...503..409H} is retained, to allow users to perform this type of analysis if desired.

\section{Examples of POSSUM Data}
\label{sec_examples}

POSSUM has accumulated a significant volume of data within its first year of routine survey operations. In this section, we present examples that showcase the transformational capabilities of POSSUM's primary outputs across both extragalactic and Galactic observing footprints.

\subsection{The RM Grid}
\label{subsec:rmgrid}

\begin{figure*}
    \centering
    \includegraphics[width=0.90\textwidth]{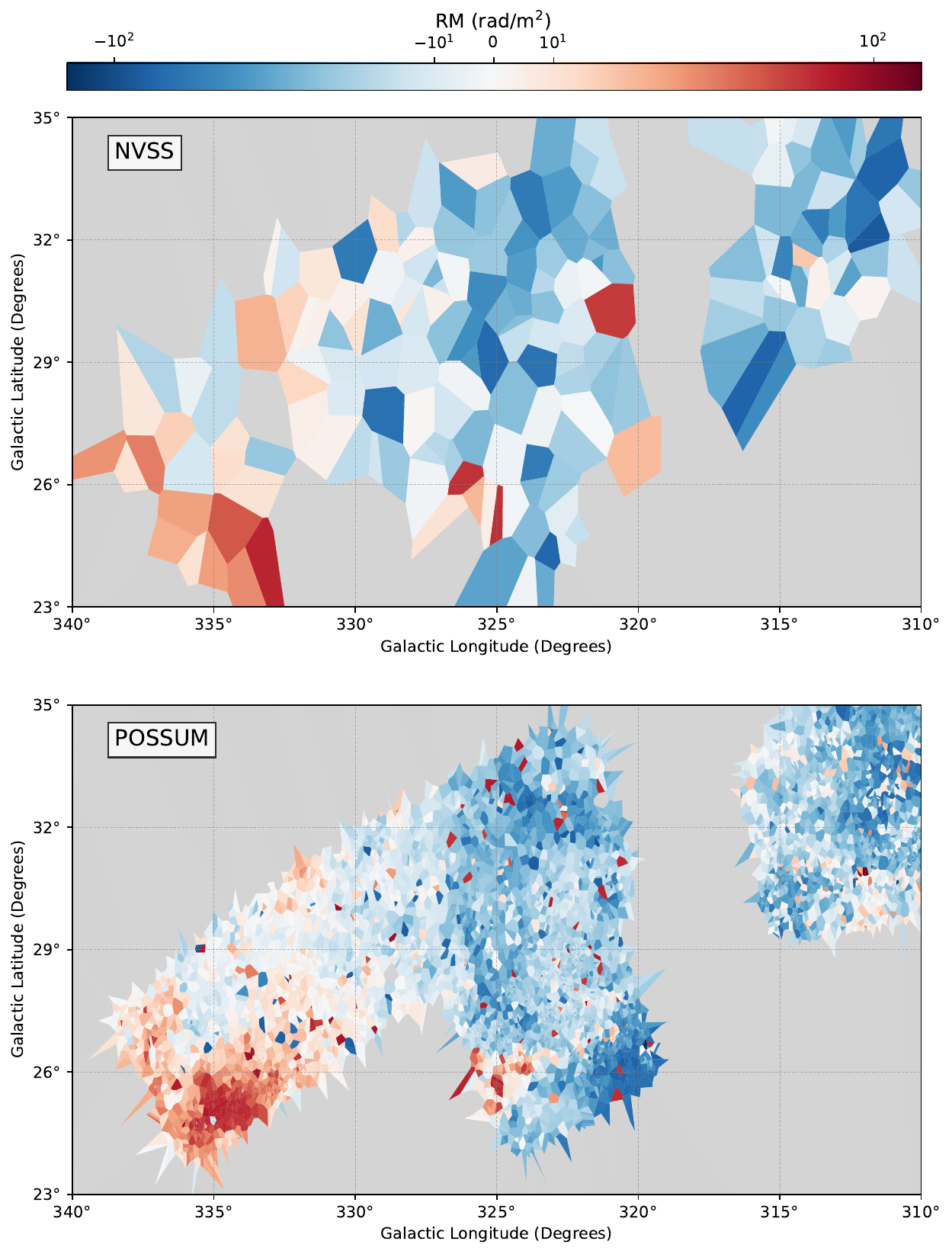}
    \caption{An illustration of the dramatic improvement in RM grid density and precision between the NVSS \citep[top;][]{2009ApJ...702.1230T} and POSSUM (bottom) band 1 surveys, focusing on mid-northern Galactic latitudes where the surveys currently overlap. POSSUM provides approximately 35 times higher RM grid density and 10 times better RM uncertainties, significantly enhancing the visibility and detail of coherent structures therein. The RM grids are shown using nearest-neighbour interpolation, where each cell's colour represents the RM of an individual linearly polarised source, normalized using \texttt{Matplotlib}'s \texttt{SymLogNorm}. The NVSS coverage has been masked to align with POSSUM's survey progress, including only NVSS sources within 1.5 degrees of a POSSUM source. Gray regions indicate areas not yet observed by POSSUM. The square tile boundaries represent ASKAP footprints, each covering approximately $5 \times 6$ degrees with 36 formed beams. The spiked polygons at the periphery are artifacts from the interpolation method used.}
    \label{fig:POSSUM_RMs_survey_overview_paper_midGalactic}
\end{figure*}

\begin{figure*}[ht]
    \centering
    \includegraphics[width=0.90\textwidth]{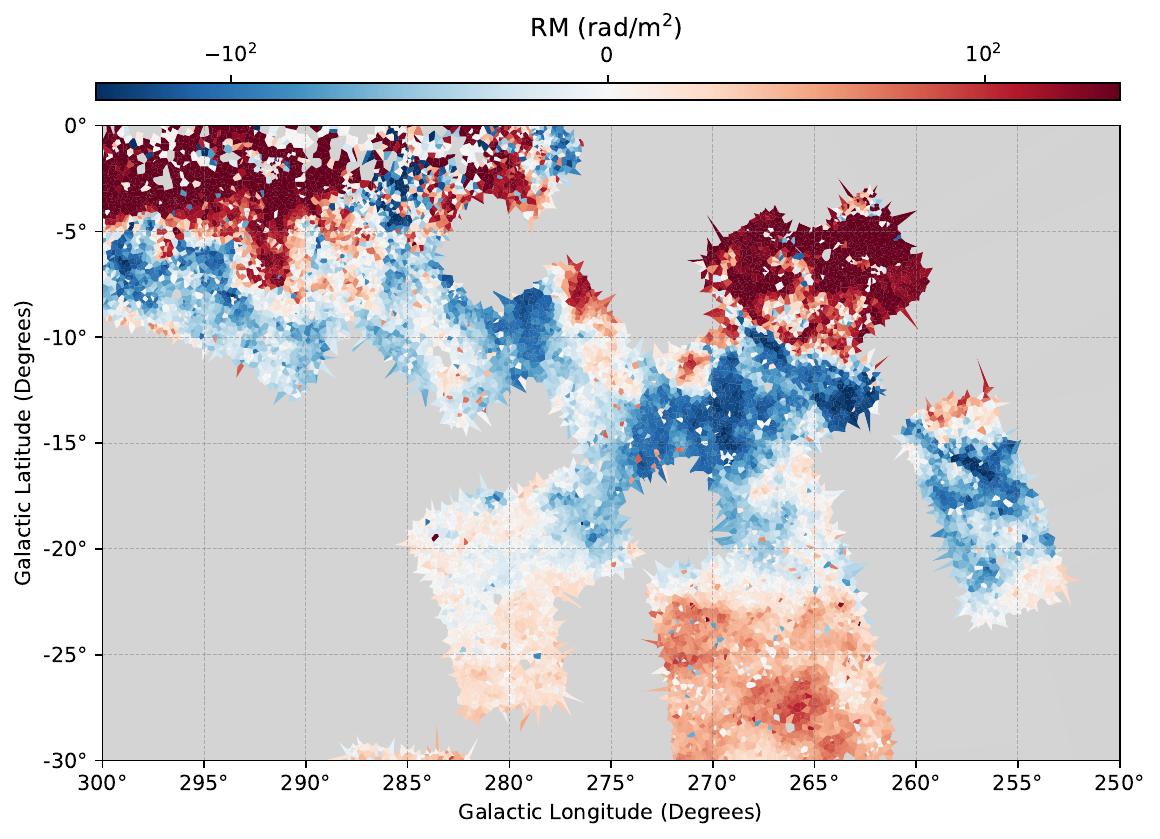}
    \caption{Similar to the bottom panel of Figure \ref{fig:POSSUM_RMs_survey_overview_paper_midGalactic} but focusing on a larger region extending south from the Galactic plane, this figure illustrates the rich variety of structures revealed by the POSSUM RM grid. These structures include small-scale filaments and ``interfaces'' where the RMs suddenly change value or sign, with the latter indicating reversals in the electron-density-weighted mean line-of-sight magnetic field direction. Closer to the Galactic plane, the absolute RM values tend to increase, while the RM grid density decreases. This reduction in grid density is due to higher noise levels in the maps of these regions, generally caused by bright diffuse Galactic ISM emission and artifacts resulting from the current absence of single-dish (zero-$uv$-spacing) data.}
        \label{fig:POSSUM_RMs_survey_overview_paper}
\end{figure*}

The dramatic improvement in RM grid capability brought about by POSSUM is vividly illustrated in Figures  \ref{fig:POSSUM_RMs_survey_overview_paper_midGalactic}, \ref{fig:POSSUM_RMs_survey_overview_paper}, and \ref{fig:fornax_RMs_comparo} \citep[see also][]{2024AJ....167..226V}. Figure \ref{fig:POSSUM_RMs_survey_overview_paper_midGalactic} highlights a region centred on a Galactic latitude of $+29^{\circ}$, where the overlapping NVSS and (incomplete) POSSUM RM grids are compared. The POSSUM grid reveals coherent and subtle RM structures not visible in the NVSS data, along with distinct outlier RMs that markedly deviate in magnitude and sign from the surrounding smoother structures. A wider view is provided in Figure \ref{fig:POSSUM_RMs_survey_overview_paper}, showing a wide range of coherent structures on various scales and an overall variation in RM magnitude and grid density with Galactic latitude.

In Figure \ref{fig:fornax_RMs_comparo}, we compare RMs measured by NVSS and POSSUM within the virial radius of the Fornax cluster. This comparison highlights the significant improvements in RM densities and uncertainties offered by POSSUM. Notably, POSSUM's ability to probe extragalactic degree-scale objects is greatly enhanced compared to previous surveys, with a mean RM separation of only $\sim10$ arcminutes. These refined measurements allowed \citet{Anderson2021} to identify a massive, previously undetected reservoir of warm plasma in the Fornax cluster.

\subsection{Extragalactic Radio Sources}

\begin{figure*}
	\centering
	\includegraphics[width=0.9\textwidth]{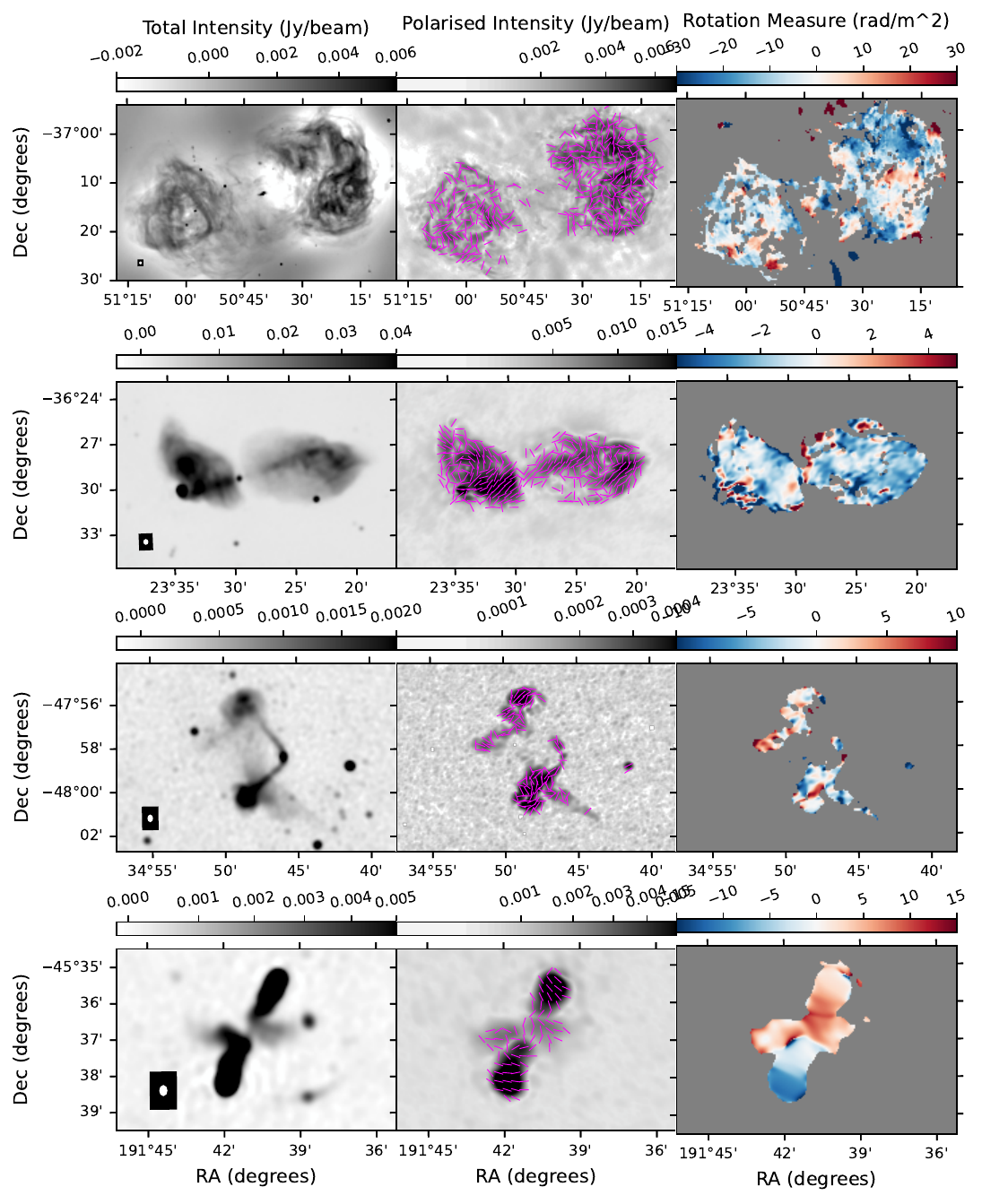}
	\caption{POSSUM band 1 images of selected radio galaxies, illustrating POSSUM's remarkable combination of resolution and sensitivity to large-scale emission. Each row features a different object, decreasing in angular size from top to bottom. POSSUM is expected to map the polarisation across the lobes of more than 100, 1,000, and 10,000 radio galaxies (respectively) of similar sizes to those shown in rows 2, 3, and 4, significantly expanding the number of galaxies we can map in detail. The left, middle, and right columns display Stokes $I$, polarised intensity, and RM maps, respectively, at POSSUM's native 20" resolution. Magenta markers on the polarised intensity maps indicate the orientation of the sky-projected magnetic field, derived from polarisation angle and RM. The synthesised beam is shown in the lower left of the Stokes $I$ plots, overlaid on a $1\times1$ arcminute box for clarity and scale. Notably, while Fornax A exceeds ASKAP's maximum recoverable scale in Stokes $I$ (resulting in a negative bowl effect), Stokes $Q$ and $U$ reveal smaller-scale structures, enabling accurate recovery of polarised intensity, RM, and angle maps (cf. \citealp{Anderson2018}). With its broad scale sensitivity, excellent polarimetric precision, and minimal depolarisation compared to lower-frequency surveys, POSSUM is ideal for RM and polarisation mapping, and complements surveys at other wavelengths.}
	\label{fig:rgs}
\end{figure*}

POSSUM's combination of a broad range between maximum recoverable angular scale and observing resolution, excellent polarimetric purity, and high RM precision, all within a frequency band where typical sources remain highly polarised, makes it an exceptionally powerful tool for producing detailed polarisation maps across a wide range of radio galaxies and other resolved objects. This capability is illustrated in Figure ~\ref{fig:rgs}, which presents polarisation maps of several radio galaxies spanning a range of brightness and angular scales in POSSUM's band 1.

One of the objects shown is Fornax A, which exceeds ASKAP's maximum recoverable scale in Stokes $I$, leading to a negative bowl effect. However, the smaller-scale structure in Stokes $Q$ and $U$ allows for accurate recovery of polarised intensity, RM, and angle maps. Remarkably, the quality of these maps is comparable to those produced by \citet{Anderson2018} using over 100 hours of targeted Australia Telescope Compact Array observations, whereas POSSUM achieves this with just 10 hours of integration in routine survey mode with automatic calibration and imaging. Intricate RM and polarisation structures are visible in Fornax A and the other galaxies, even at sub-beam scales, due to the vector nature of polarisation. Such structures can be used to infer complex internal dynamics within radio lobes (e.g., \citealp{Anderson2018}) and interactions with their environments \citep[e.g.,][]{Guidetti2010, 2011MNRAS.413.2525G,2012MNRAS.423.1335G}.

In addition to nearby, highly resolved radio galaxies, POSSUM will image many fainter and less extended sources. Based on detections from EMU \citep{Gupta2024} and LoTSS \citep{Williams2019}, and assuming an average polarisation fraction of 25\% for resolved radio lobes with a spectral index of -0.8, we expect POSSUM to map polarisation across more than 100, 1000, and 10,000 objects similar in scale and brightness to those shown in rows 2, 3, and 4, respectively, of Figure ~\ref{fig:rgs}. Though less prominent than the brighter 3C sources, these galaxies are still sufficiently bright and well-resolved to provide detailed polarisation and RM maps. This will significantly enhance our ability to conduct population studies of radio galaxies and other astrophysical objects.

\subsection{Galactic Objects and ISM-Dominated Fields}

\begin{figure*}
    \centering
    \includegraphics[width=1\textwidth]{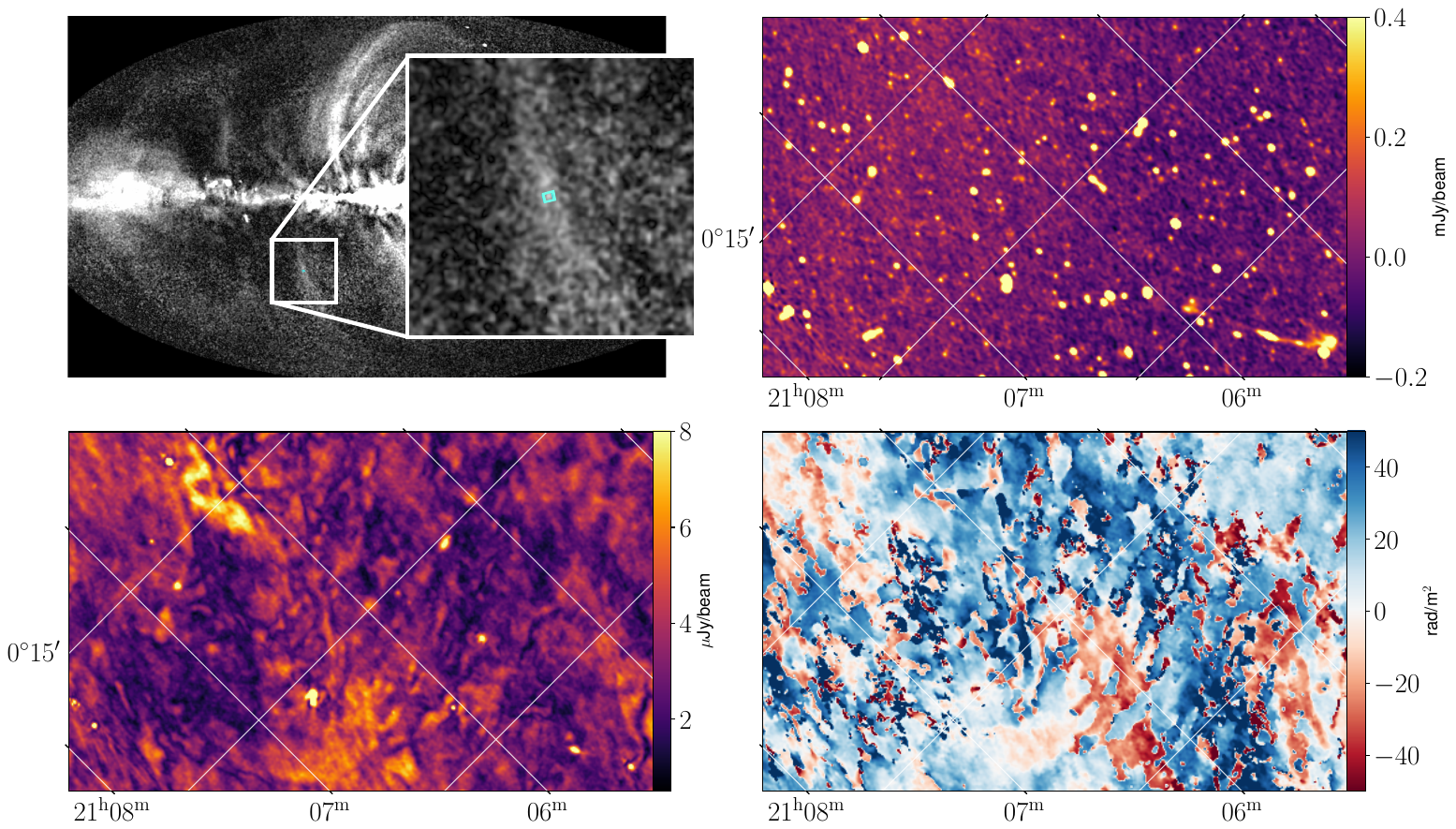}
    \caption{Part of a field centred at $(\ell,b)=(49.75^\circ,-29.5^\circ)$, demonstrating POSSUM's excellent sensitivity to Galactic ISM emission structures. The top left panel shows the all-sky Wilkinson Microwave Anisotropy Probe \citep[WMAP;][]{2003ApJ...583....1B} linearly polarised intensity image for context, with an inset highlighting a region of substantial polarised emission. A cyan box indicates the area shown in the other panels. The top right panel provides a zoomed-in view of this area, displaying the ASKAP Stokes $I$ map from the EMU survey. The bottom left panel shows the ASKAP linearly polarised intensity measured by the POSSUM survey, while the bottom right panel presents the peak RM derived from RM synthesis using POSSUM data. These observations reveal significant Galactic diffuse emission, sufficiently bright to reveal notable and coherent RM structures in this region. Unlike Figures \ref{fig:POSSUM_RMs_survey_overview_paper_midGalactic} and \ref{fig:POSSUM_RMs_survey_overview_paper}, where coherent RM structures are revealed by extragalactic polarised background sources, here the RM structure is both generated and illuminated by the Galactic ISM.}
    \label{fig:galactic}
\end{figure*}

POSSUM provides excellent surface brightness sensitivity to discrete Galactic objects and resolved Galactic ISM emission. The polarised emission from the diffuse Galactic ISM serves as a key probe of magnetic fields and thermal gas. POSSUM can detect diffuse polarised emission over much of the sky (see Figure \ref{fig:galactic}), enabling new studies of the Galactic ISM, helping to disentangle Galactic emission from extragalactic polarisation and RM studies.

\begin{figure*}
    \centering
    \includegraphics[width=1\textwidth]{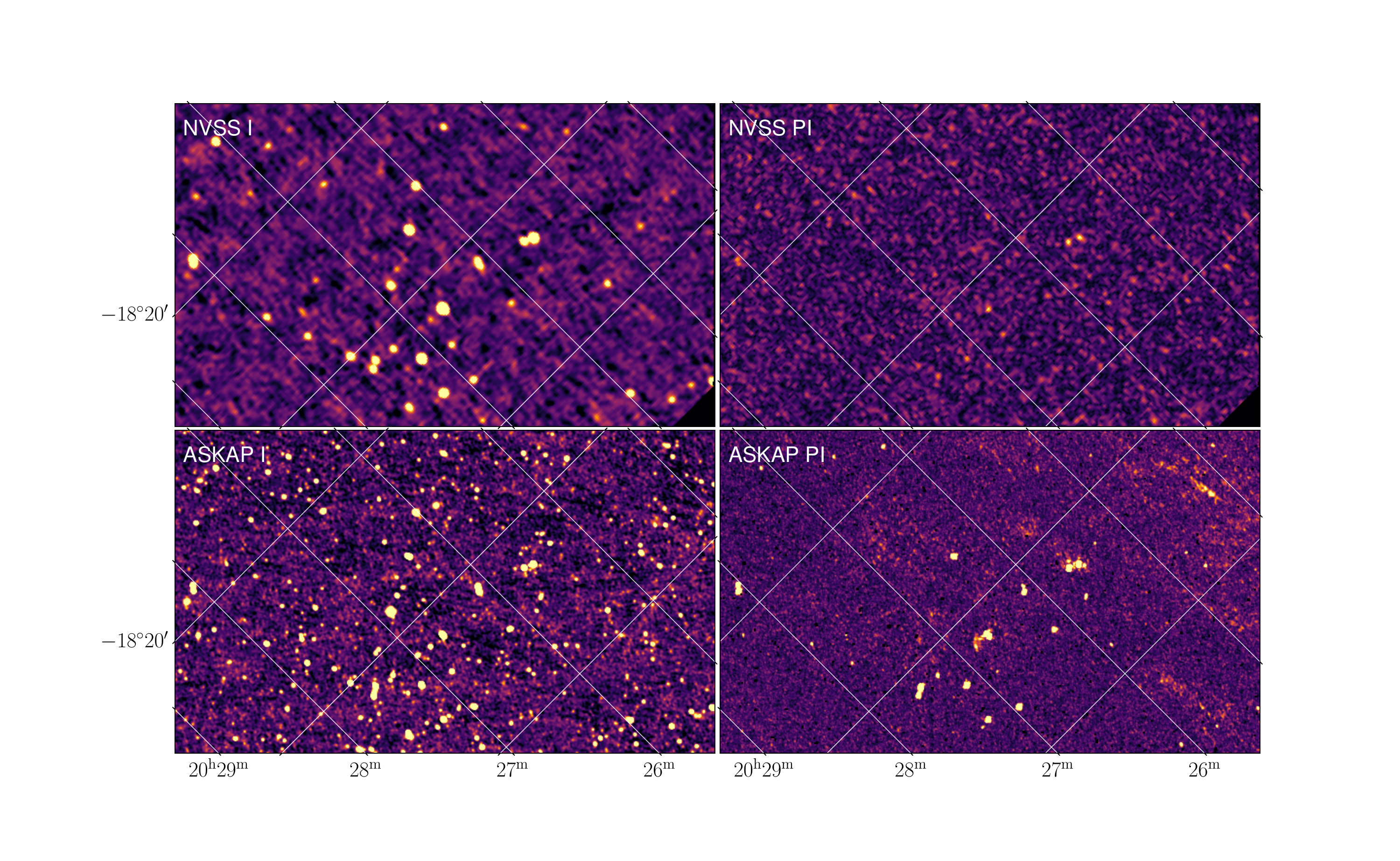}
    \caption{Comparison of NVSS \citep{1998AJ....115.1693C} and ASKAP EMU/POSSUM maps in a typical extragalactic field. The top left panel shows the NVSS Stokes $I$ map, while the top right panel presents the corresponding NVSS linearly polarised intensity. The bottom left panel displays the ASKAP Stokes $I$ map from the EMU survey, and the bottom right panel shows the linearly polarised intensity derived from RM synthesis using POSSUM data. Notably, POSSUM reveals approximately 27 polarised detections in this region, whereas the NVSS data shows no catalogued polarised sources. The POSSUM polarisation map also reveals faint ISM emission structure, even in a largely empty field, a feature frequently observed in POSSUM data but not seen in NVSS.}
    \label{fig:nvss}
\end{figure*}

\section{SYNERGIES WITH OTHER SURVEYS}
\label{sec_others}

\subsection{Other ASKAP Surveys}
\label{sec:other_ASKAP}

\subsubsection{The Evolutionary Map of the Universe (EMU)}
\label{sec:emu}
The Evolutionary Map of the Universe survey \citep[EMU;][]{2011PASA...28..215N,2021PASA...38...46N} 
focuses on analysis of the total intensity from the same ASKAP band-1 data used by POSSUM. The EMU survey uses these data to produce multi-frequency synthesis images with a uniform angular resolution of $15''$, along with a catalogue expected to total about 24 million radio sources. EMU also produces higher resolution images using uniform weighting, with FWHM $7'' - 10''$, but with roughly double the noise level of the $15''$ images. Key goals for EMU include tracing the evolution of star-forming galaxies from $z= 3$ to the present day, tracing the evolution of massive black holes throughout the history of the Universe, exploring the large-scale structure and constraining cosmological parameters, testing fundamental physics, creating a wide-field atlas of Galactic continuum emission to address areas such as star formation, supernovae, and Galactic structure, and searching for new classes of objects \cite[e.g.,][]{2021PASA...38....3N}.  

As noted in \S\ref{1d_pipeline} and \S\ref{sec:POSSUM_catalogue}, the EMU catalogue provides the positions for the analyses carried out by the POSSUM 1D polarisation pipeline. Although only $\approx 5\%$ of EMU sources will yield reliable individual RMs,\footnote{\citet{2024AJ....167..226V} find $\approx 50$\,RM\,deg$^{-2}$ at high latitude in their combined-band analysis, while unpublished EMU survey results find $\approx 1100$ sources deg$^{-2}$ at high latitude.} upper limits to polarization will be available for the remainder, and
polarisation properties of fainter sources can be studied by stacking
\citep[cf.,][]{2014ApJ...787...99S}. The EMU collaboration is devoting substantial effort to cross-matching sources with surveys in other wavebands, and hence finding distance or redshift estimates: this will be immensely valuable for POSSUM. 

The EMU MFS total intensity images have significantly better image fidelity than the band-averaged POSSUM total intensity cubes, due to a much lower CLEAN threshold and improved $uv$ coverage from MFS, as well as slightly higher angular resolution ($15''$ for EMU vs $20''$ for POSSUM). The POSSUM archive includes versions of the MFS images smoothed to POSSUM resolution, from which images of fractional polarization can be derived. For resolved sources, additional structural details visible in the EMU high-resolution images will play a key role in both identifying interesting targets and in interpreting the POSSUM polarisation maps.

\subsubsection{The Rapid ASKAP Continuum Survey (RACS)} \label{sec_racs}

The Rapid ASKAP Continuum Survey \citep[RACS;][]{McConnell2020} is an observatory-led project that has imaged the entire sky visible from ASKAP ($-90^\circ<\delta<+49^\circ$). RACS is conducted in each of ASKAP's available bands, which are centred on 887.5/943.5~MHz (RACS-low), 1367.5~MHz (RACS-mid), and 1655.5~MHz (RACS-high). These observations serve a dual purpose: they provide a sky model for enhanced calibration of ASKAP continuum surveys such as EMU and POSSUM, and also function as a shallow all-sky survey in their own right.

RACS and SPICE-RACS~\citep[the linear polarisation catalogue derived from RACS,][]{2023PASA...40...40T} offer several synergies with POSSUM. First, SPICE-RACS provides an initial characterization of relatively bright polarised sources, establishing a preliminary RM grid with lower sky density compared to POSSUM, and facilitating cross-comparison and validation of POSSUM's results. Additionally, RACS observations span all three ASKAP bands, enabling ultra-broadband analysis of bright polarised sources. Finally, RACS is intended to be repeated across multiple epochs in parallel with the POSSUM survey, allowing for the study of variability in polarised sources and their properties through cross-comparison of these epochs.

\subsubsection{H\,{\sc i} Surveys (WALLABY, GASKAP, FLASH)}
POSSUM band-2 observations are commensal with the Widefield ASKAP L-band Legacy All-sky Blind surveY \citep[WALLABY;][]{2020Ap&SS.365..118K}, which aims to detect neutral hydrogen in hundreds of thousands of nearby galaxies ($z \lesssim 0.08$), with 4\,km\,s$^{-1}$ velocity resolution. Tens of thousands of these detections will be spatially resolved, enabling comparison with polarised synchrotron emission mapped by POSSUM to characterise systematic patterns of disk polarisation in all classes of star-forming galaxies.

The Galactic ASKAP (GASKAP) surveys \citep{Dickey2013,2022PASA...39....5P} are mapping the H\,\textsc{i} distribution in the Milky Way and the Magellanic system, as well as OH in selected regions, with velocity resolution as fine as 0.2\,km\,s$^{-1}$. These surveys complement POSSUM by tracing the (nominally) neutral ISM, which is believed to contribute a significant fraction of the Galactic electron column density \citep[about 10\% in the model of][]{McKee1977}, and hence the RM -- direct evidence for Faraday rotation in the neutral ISM was found by \citet{VanEck2019}. Moreover, cold H{\sc i} traces the plane-of-sky magnetic field via narrow (width $\approx 0.1\,{\rm pc}$) H\textsc{i} filaments in the Milky Way
\citep[e.g.,][]{2006ApJ...652.1339M,2014ApJ...789...82C}, 
and potentially also the $\approx 10\,{\rm pc}$-wide filaments in the SMC \citep{2023MNRAS.521...60M}. ASKAP's high resolution will allow detection of such Galactic structures out to distances of $\sim 1$\,kpc (several times further than previous surveys), as well as in the Magellanic system.

The First Large Absorption Survey in H\,\textsc{i} \citep[FLASH;][]{Allison2022} is expected to detect thousands of cold neutral 21-cm H\,\textsc{i} absorbers in both intervening and associated systems over the redshift range $0.4<z<1.0$ across the sky south of +40$^\circ$ declination, with a frequency resolution of 18.5 kHz (equivalent to 5.5--7.8 km\,s$^{-1}$ depending on redshift).
As FLASH is an untargeted survey, it is expected to provide a dust-unbiased view of the H\,\textsc{i} absorber population, potentially revealing fainter systems than previously possible \citep[see, e.g.,][who assembled the largest sample to date]{Curran2021}.

As noted in \S\ref{sec:magellanic}, POSSUM polarisation information on the bright background continuum sources against which FLASH detects H\,\textsc{i} absorption will be an important probe of magnetic fields in normal star-forming galaxies at intermediate redshift. In early observations with ASKAP, \citet{Mahony2022} have already detected intervening 21-cm absorption towards a radio lobe of PKS~B0409--75, and used archival  polarisation data to estimate the magnetic field of the intervening galaxy to be $\sim 14\,\mu$G, broadly consistent with field strengths of nearby galaxies. POSSUM will provide much more sensitive and precise polarisation data than that used by \citet{Mahony2022}. Such work will largely be statistical since, unlike H\,{\sc i} absorption, Faraday rotation is an integral of all contributions along the line of sight. Hence, to establish a clear association between excess RM and \HI\ absorption, we need control samples, which will be provided by POSSUM polarisation information on FLASH non-detections. POSSUM/FLASH results will also quantify the population of intervening RM systems, which is essential for studies of intergalactic magnetic fields. FLASH will also probe the neutral gas within the active galaxy itself, as discussed  \S\ref{sec:AGN}.

\subsubsection{VAST}
The ASKAP Variables and Slow Transients Survey aims to observe the sky for highly variable and transient sources on timescales as short as 5 seconds \citep{murphy13}. Synergies between POSSUM efforts and transient searches have already proved to be successful such as a recent serendipitous discovery of a highly polarised point source \citep{kaplan19} and ASKAP detection of periodic and elliptically polarised radio pulses from UV Ceti \citep{pritchard21}. There has also been a survey of circularly polarised radio stars using ASKAP \citep[e.g.][]{zic19}. Further synergies between POSSUM and VAST will increase the sample size of transients and Galactic sources with polarised emission as well as enable serendipitous discoveries.

\subsection{Other Interferometric Polarisation Surveys}
\label{sec_other_pol}

Table~\ref{tab:polsurveys} lists the parameters of current large-area polarisation surveys, most of which will be complementary with POSSUM. We summarize these surveys and their relevance in the following subsections.

\subsubsection{The Very Large Array Sky Survey (VLASS)}
The Very Large Array Sky Survey \citep[VLASS;][]{Lacy2020} is a three-epoch, 2--4~GHz survey covering $\delta > -40\degr$, thus covering 64\% of the POSSUM area. 
VLASS has significantly higher angular resolution but lower sensitivity than POSSUM, especially for extended sources, as the $uv$ coverage is relatively sparse; thus source components well resolved by POSSUM will usually not be detectable in VLASS.
The synergy of VLASS with POSSUM is to reveal the spatial structure of marginally resolved or unresolved POSSUM sources, find accurate spectral indices of compact components, detect variability between VLASS epochs, and, most importantly, to extend the overall $\lambda^2$ coverage in polarisation by providing combined (POSSUM+VLASS) frequency coverage from 0.8 to 4\,GHz. This range is large enough to resolve the kind of substructure found in the FDF of Faraday-thick sources such as powerful quasars \citep{Garrington1988} and the cores of rich clusters of galaxies \citep{Anderson2016}, often caused by distinct regions with very different Faraday dispersions that give a highly non-Gaussian FDF, or, equivalently, gradual depolarization over a broad range of wavelengths. Even sources seen by POSSUM to exhibit moderate depolarisation will benefit from the complementary information provided by VLASS at higher frequencies, allowing characterisation of the intrinsic fractional polarisation and thus the underlying physical conditions \citep[e.g.,][]{2017MNRAS.469.4034O}. We also expect numerous Faraday-thick sources only detectable in VLASS, and conversely, many sources that appear Faraday simple in VLASS but show complexity with POSSUM's finer RM resolution. In combination, the two surveys will give a much more complete picture of the Faraday properties of the polarized radio source population.

Observations for the third epoch of VLASS will be completed in late 2024. Stokes $I$ images from completed epochs are already available; polarisation products are scheduled for release from 2026, culminating in $\sim$2032--2033 with the release of combined-epoch data at full frequency resolution, cf.\ footnote $j$ to Table~\ref{tab:polsurveys}.\footnote{See {\tt science.nrao.edu/vlass/vlass-data/basic-data-products}.} 
This latter dataset is important as it will allow detection of the very large RMs ($\sim 10^4 {\rm \, rad\,m^{-2}}$) found in exceptionally dense environments such as the Galactic centre. Although in principle these are detectable by band-2 POSSUM, i.e., RM $< \phi_{\rm max}$, in practice even a fractional spread in RM of 1\% would exceed $\phi_{\rm max-scale}$  and hence the source would be heavily depolarized, whereas $\phi_{\rm max-scale}$ is eight times larger for VLASS. 

\subsubsection{Aperture Tile in Focus (Apertif)}
The Aperture Tile in Focus \citep[Apertif;][]{vancapellen22} was a PAF system on the Westerbork Synthesis Radio Telescope (WSRT), operating over 1292--1442\,MHz (after RFI excision) and serving as a northern-hemisphere counterpart to ASKAP, with similar science goals to the various ASKAP surveys. Due to the larger WSRT dishes, the field of view was $\sim$8\,deg$^2$ compared to ASKAP's $\sim$30\,deg$^2$. From 2019 to 2022, an imaging survey covered about 360 fields, mostly with single 11.5-hour tracks, giving sensitivity comparable to POSSUM. Twenty fields were observed with up to ten tracks each, to produce a medium-deep survey. The fields are at $\delta \ge +27^\circ$, except for three near the North Galactic Pole at $\delta \approx +25^\circ$. A preliminary reduction of the first year's data has been presented by \citet{2022A&A...667A..38A}.

The Apertif and POSSUM surveys have no sky overlap. Apart from providing about a 15\% increase in the solid angle mapped in linear polarisation with quality similar to POSSUM band-2, the main synergy provided by Apertif is its access to the polarisation of sub-mJy sources in its medium-deep fields, which target regions with extensive multi-band coverage including the {\it Herschel}-ATLAS NGP field \citep{2017ApJS..233...26S}.

\subsubsection{Low-freqeuncy surveys (LoTSS and POGS)}
\label{sec_low_freq}

Low-frequency polarimetric observations are being conducted by the LOFAR Two-Metre Sky Survey \citep[LoTSS;][]{shimwell19,shimwell22}, covering the northern hemisphere, and the GaLactic and Extragalactic All-sky MWA-eXtended survey \citep[GLEAM-X;][]{2022PASA...39...35H}, which covers $-90\degr < \delta < +30\degr$. LoTSS has released polarisation products, including an RM grid \citep{2023MNRAS.519.5723O}, while MWA-based data are available through the Polarised GLEAM Survey \citep[POGS;][]{2018PASA...35...43R,2020PASA...37...29R} and POGS-X (Zhang et al., in prep). LoTSS polarisation data offer angular resolution and signal-to-noise comparable to POSSUM (assuming a spectral index of $-0.7$), whereas GLEAM-X is shallower with a resolution of $\sim 60''$.

At these frequencies, strong Faraday depolarisation means that the vast majority of extragalactic sources are not detected in linear polarisation; POGS detected only 0.16\% of the sources seen by GLEAM in total intensity, despite GLEAM (but not POGS) being confusion-limited. The final LoTSS release will have $6''$ resolution, which is expected to resolve many of the polarised sources, reducing beam depolarisation and thereby increasing the source detection fraction. The new iLoTSS project\footnote{\tt https://lofar-surveys.org/ilotss.html} will use LOFAR international baselines to push the resolution to $0\farcs 3$.

Nevertheless, for sources that are detected, these surveys benefit from very large $\Delta \lambda^2$, providing extremely good RM resolution. This precision enables the study of magnetic fields in low gas density environments, such as the WHIM expected to permeate cosmic web filaments \citep[e.g.,][]{2022MNRAS.512..945C,2020A&A...638A..48S}, which complements POSSUM's capabilities probing denser environments. 

The combined study of polarised sources seen in both surveys using the resulting very large $\lambda^2$ coverage will provide insights into the underlying mechanisms for the observed depolarisation \citep[e.g.][]{2018MNRAS.475.4263O}. Such comparisons can only be made directly over the $\approx$ 1310\,deg$^2$ (3.2\% of the sky) overlap between the two surveys, but this may usefully inform interpretation of the full results of both surveys. 

Small scale features are often apparent in low-frequency diffuse Galactic polarised emission, even when smooth Stokes $I$ emission structures have been resolved out \citep{Wieringa1993}. An RM gradient in front of a smooth polarised background induces structure on small angular scales.  The effect is wavelength dependent;  increasingly small angular scale structures are observed at longer wavelengths. LoTSS is releasing $(Q,U)$ cubes at $3'$ resolution, in which Faraday synthesis detects emission over most of the sky \citep{Erceg2022,Erceg2024b,Erceg2024}. Interpretation of the observed diffuse polarised emission requires the largest possible frequency coverage because Faraday rotation and emission are mixed along the line of sight, producing broad structure in Faraday depth spectra. An analysis combining LoTSS and POSSUM$+$PEGASUS diffuse polarisation maps (see \S\ref{sec_ISM} and \S\ref{sec_single_dish_surveys}) will be extremely valuable, even with their limited sky overlap. Similar products can be extracted from the GLEAM data in the southern hemisphere but are not included in POGS(-X), which focus on discrete sources. A region covering 660\,deg$^2$ (1.6\% of the sky) at high latitude was analysed by \protect\citet{Lenc_2016}, but currently there are no plans to extend this to the full survey.

\subsubsection{Galactic Plane Purveys: THOR and MMGPS}

Studies of the Galactic plane are important science drivers for POSSUM (\S\ref{sec_ISM}), and many other surveys across the electromagnetic spectrum provide useful synergies. Here, we focus on two  surveys of Faraday rotation in the Galactic plane that directly complement POSSUM.

The HI/OH Recombination line survey of the inner Milky Way \citep[THOR;][]{beuther16,2022ApJ...939...92S} was a VLA C-configuration project
that included 1--2\,GHz continuum  polarisation data. Its extension to cover the Galactic centre, THOR-GC \citep[e.g.,][]{2024arXiv240513183W} includes D-configuration data to improve sensitivity to large angular scales.  

As a polarisation survey, THOR's advantages and disadvantages are similar to those of VLASS, but its parameters are a better match to POSSUM, allowing near continuous extension of frequency coverage up to 2 GHz, apart from a large gap around 1520\,MHz due to GPS satellites. THOR provides a maximum Faraday depth of $2.5 \times 10^4\ \rm rad\ m^{-2}$ using the top half of the band at full spectral resolution (2 MHz). This has already allowed the discovery of a sharp peak in RM (up to 4219\,rad\,m$^{-2}$) near the tangent point of the Sagittarius spiral arm \citep{shanahan2019}. THOR-GC is entirely within the POSSUM area. 

The Max Planck Institute for Radio Astronomy (MPIfR) leads the MPIfR-MeerKAT Galactic Plane Survey \citep[MMGPS;][]{2023MNRAS.524.1291P}. While the foremost objective of MMGPS is a pulsar search, it also provides 
spectral line (CH, H\,{\sc i}) and continuum polarisation data. The MMGPS is divided into five sub-surveys: the sky coverage and parameters of the three most relevant to POSSUM are listed in Table~\ref{tab:polsurveys}. In their overlap frequency range,
POSSUM is significantly more sensitive than MMGPS, especially for diffuse emission due to its lower resolution and denser short-baseline coverage, but MMGPS substantially extends the frequency coverage, giving advantages similar to those described for THOR and VLASS, although MMGPS will be the preferred partner along the plane at 2-3\,GHz since it greatly improves on VLASS in both noise and sensitivity to extended structure.

Both THOR-GC and MMGPS overlap POSSUM in both sky and frequency coverage, giving opportunities for both data verification and science, since high-quality RM surveys at different epochs can reveal variability of RM in ultracompact sources \citep[e.g.,][]{2024arXiv240513183W}, as well as variations which probe changing structure in parsec-scale AGN \citep{2019MNRAS.485.3600A}.

\subsection{Single-Dish Polarisation Surveys}
\label{sec_single_dish_surveys}

Modern single-dish polarisation surveys are motivated both by the need for foreground correction of cosmic microwave background 
(CMB) polarisation \citep[e.g.,][]{2019MNRAS.489.2330C,2018MNRAS.480.3224J,2023MNRAS.519.3383R}, and by the exploration of the Faraday structure of diffuse Galactic synchrotron emission. This effort is dominated by the Galactic Magneto-Ionic Medium Survey \citep[GMIMS;][]{2019AJ....158...44W, 2021AJ....162...35W}, which aims to observe the entire sky over 300--1800\,MHz by combining six component surveys. 

Thanks to ASKAP's excellent short-spacing coverage, POSSUM maps are sensitive to structures up to $0\fdg 5$ across, although these scales are significantly undersampled. Such structures occur along the Galactic plane and in large extragalactic objects like Centaurus~A \citep[e.g.,][]{Anderson2018b} and the Magellanic Clouds, creating substantial artifacts in the images, including large negative `bowls' around extended sources. Similarly, the diffuse polarised emission in POSSUM images is often undersampled, especially at the short-wavelength end of the band where the fragmentation of diffuse polarisation described at the end of \S\ref{sec_low_freq} is less effective. The long-term plan is therefore to remove such artifacts by combining ASKAP continuum survey data (both POSSUM and EMU) with single-dish surveys. There is excellent overlap in the Fourier plane between long-track ASKAP observations and those from the 64-m Murriyang Radio Telescope at Parkes, for example, as required for accurate combination. 

POSSUM will be combined with different single dish surveys for each of its band-1 and band-2 components. The frequency range of POSSUM band 2 is almost entirely covered by the Southern Twenty-centimetre All-sky Polarisation Survey (STAPS), conducted with Murriyang \citep{Raycheva2024,2025A&A...694A.169S}. STAPS is the high-band South component of GMIMS: it covers $-90\degr < \delta <0^\circ$ over 1328--1768\,MHz, with a frequency resolution of 1\,MHz. The missing coverage at $\delta > 0\degr$ could be supplied by GMIMS high-band North \citep{2021AJ....162...35W}, although there is little diffuse Galactic emission in this part of POSSUM. STAPS was observed commensally with the 2.3\,GHz S-PASS survey \citep{2019MNRAS.489.2330C}, using an off-axis feed. To cover band 1, the POSSUM-EMU-GMIMS All Stokes UWL Survey (PEGASUS, Carretti et al., Parkes project ID P1123) is currently underway on Murriyang. PEGASUS will cover $-90\degr < \delta < +20^\circ$ over 700--1440\,MHz, with a final frequency resolution of 0.5\,MHz, with a scan strategy similar to S-PASS and STAPS. The frequency overlap with STAPS will allow cross-calibration, although the new PEGASUS data are superior due to improved RFI rejection and an on-axis feed.  

At the low angular resolution of the single-dish maps ($20'$--$60'$), the surface brightness sensitivity is significantly higher than in the POSSUM images, which will not detect faint Galactic emission at high latitude. Although it may seem excessive to map the entire hemisphere when coverage is primarily required over some 7000\,deg$^2$ along the Galactic plane and in a few other specific regions, (a) it is not practical to scan restricted regions of the sky while preserving the large-scale structure, and (b) PEGASUS also functions as the mid-band South component of the broader GMIMS survey project.

PEGASUS observations are expected to be complete in 2025, three years ahead of POSSUM. Unlike POSSUM, the PEGASUS data cannot be fully reduced until observations are completed, since adjacent scans are observed in different observing runs and the final reduction relies on scan crossings to solve for residual baseline offsets (destriping). The first images of the diffuse polarisation from combined POSSUM and PEGASUS data will be created as enhanced products once PEGASUS reduction is complete.   

\subsection{POSSUM Exploitation Projects: QUOCKA and QUOLL}

Like other large surveys, POSSUM will often serve as a finder survey for objects deserving more detailed follow-up.

An early example is the QU Observations at Cm Wavelength with Km baselines using ATCA project (QUOCKA; Heald et al., in prep.), aimed at clarifying the magnetoionic structure of AGN jets and lobes and their surroundings. QUOCKA targets 537 radio galaxies with reliably detected ASKAP polarisation and brightness suitable for detection with the Australia Telescope Compact Array, selected from 13 ASKAP Early Science fields. 

QUOCKA observations cover 1.1--8.5,GHz, spanning a factor of 60 in $\lambda^2$, thereby dramatically improving sensitivity to Faraday complexity generated by complex, strongly magnetised environments such as radio galaxies. QUOCKA also provides a sensitive probe of inner jet structure, measuring circular polarisation with minimal residual leakage. Each source was observed for about 30 minutes in each band, split into six 5-minute snapshots over 12 hours to improve $uv$ coverage. After convolving all frequency planes to a common resolution, QUOCKA data typically have $10''-15''$ resolution and a broadband sensitivity of $\approx50$~$\mu$Jy~beam$^{-1}$. QUOCKA's broadband spectropolarimetry is expected to be invaluable as a polarisation reference, and was proven so during the commissioning of ASKAP's polarimetric performance, helping to identify spurious features in the ASKAP data and guiding the necessary corrections to achieve the data quality required by POSSUM.

Since many QUOCKA sources are dominated by structures unresolved by ATCA, the Australian Long Baseline Array (LBA) has been used to map a few of the brighter QUOCKA sources with the QU Observations Leveraging the LBA project (QUOLL; Kaczmarek et al., in prep.) project. The milliarcsecond resolution at both 1.4 and 10\,GHz will be sufficient to resolve sub-kpc scale structures, mapping the detailed polarised morphology similar to the MOJAVE survey \citep[e.g.,][]{2012AJ....144..105H,2005AJ....130.1389L,2018ApJS..234...12L} but at frequencies suitable for tracing depolarisation effects, and linking directly with the $QU$ spectra from ATCA.

\section{SUMMARY AND CONCLUSIONS}
\label{sec:conclusion}

The Polarisation Sky Survey of the Universe's Magnetism (POSSUM) represents a major leap forward in the study of cosmic magnetic fields. Using the advanced capabilities of ASKAP, POSSUM aims to create the most detailed maps of the magnetoionic Universe across the southern sky. This survey will generate a comprehensive Faraday rotation measure grid covering half the sky, along with $IQU$ polarisation spectra of these RM grid sources, and detailed 2D polarisation maps and 3D Stokes $IQU$ spectral cubes of both discrete objects and diffuse emission.

POSSUM's RM grid achieves a density of 30--50 RMs per square degree with a median uncertainty of approximately 1 rad m$^{-2}$, dramatically surpassing previous surveys like the NVSS RM catalogue. Similarly, POSSUM's excellent surface brightness sensitivity enhances the detection of polarised emission from the diffuse ISM and from a wide variety of discrete radio sources. This combination of high RM grid density, precise RM measurements, and superior imaging capability addresses a range of scientific themes in diverse cosmic environments, including the study of the intergalactic medium, galaxy clusters and groups, the circumgalactic and interstellar media, discrete objects in the Milky Way, nearby galaxies, active galaxies, and fundamental physics. Furthermore, the simultaneous production of a high-quality RM grid and high-quality images also provides significant benefits for contextualising and interpreting each type of measurement, particularly regarding the impact of Galactic foreground Faraday rotation. For example, diffuse ISM polarisation can help determine whether Faraday rotation enhancements seen in the RM grid are Galactic in origin (e.g., \citealp{2021MNRAS.508.3921J}). Conversely, conducting a structure function analysis on the RM grid can provide insights into whether RM variations observed toward a particular source are intrinsic to that source or influenced by intervening material (e.g., \citealp{Anderson2018}). Consequently, POSSUM serves as a comprehensive resource to advance our understanding of cosmic magnetism across different scales and locales.

Cosmic magnetism remains a crucial science driver for future observatories like the SKA, where it is one of the original five key science projects \citep{CR2004}. The core goal of SKA's Magnetism Science Working Group is to create a broadband, dense, wide-area, high-resolution RM grid with SKA1-MID \citep{Gaensler2004,Beck2004,2015aska.confE..92J,Heald2020}. POSSUM is the only pathfinder specifically dedicated to a comprehensive study of cosmic magnetism. Its unique combination of broad sky coverage, high RM-grid density, densely-sampled wavelength-squared coverage, excellent sensitivity, low RM uncertainties, and high-fidelity imaging all pave the way for large-scale magnetism studies with the SKA. Consequently, POSSUM is developing essential methods, software, and community expertise required for the SKA's ambitious magnetism science projects \citep{Heald2020}.

There are several avenues to enhance and extend POSSUM. All calibrated visibilities, holography data, and beam-forming information used to generate the leakage-corrected POSSUM data are available online via the CASDA repository. This allows for straightforward re-processing of POSSUM data with advanced calibration and imaging techniques, ranging from a single beam to the entire survey. ASKAP's unique mode of observing---where multiple formed beams overlap to cover the same sky area, and where different sky areas are observed with the same formed beams---enables the application of powerful statistical tools to further identify and correct residual leakage or systematic biases. Beyond the current survey plan, future ASKAP time allocations or mid-survey sensitivity upgrades could enable us to extend our sky coverage to +50$^{\circ}$ in declination, covering 36,100\,deg$^2$ (87.5\% of the sky). This would benefit all POSSUM science areas, but particularly those leveraging broadband radio or multi-wavelength observational coverage. Additionally, we could conduct deep-drill polarimetric observations that push well beyond the Stokes I confusion limit, or extend POSSUM's frequency coverage to the upper limit of ASKAP's high-frequency band (1.8 GHz). The scientific benefits of deep observations of a smaller area would accrue more towards studies of intrinsic source properties rather than the RM grid.\\

In conclusion, the POSSUM  collaboration is producing a comprehensive, high-quality radio polarisation dataset that addresses numerous important contemporary scientific themes and lays the groundwork for future research on cosmic magnetism with the SKA. POSSUM will answer many current questions in the field, will pave the way for groundbreaking discoveries, and will provide  significant legacy value for studying the ionised, magnetised Universe.

\begin{acknowledgement}

The authors thank the anonymous referee for their valuable feedback, which has improved both the clarity and utility of this work.

This scientific work uses data obtained from Inyarrimanha Ilgari Bundara, the CSIRO Murchison Radio-astronomy Observatory. We acknowledge the Wajarri Yamaji People as the Traditional Owners and native title holders of the Observatory site. CSIRO's ASKAP radio telescope is part of the Australia Telescope National Facility (\text{https://ror.org/05qajvd42}). Operation of ASKAP is funded by the Australian Government with support from the National Collaborative Research Infrastructure Strategy. ASKAP uses the resources of the Pawsey Supercomputing Research Centre. Establishment of ASKAP, Inyarrimanha Ilgari Bundara, the CSIRO Murchison Radio-astronomy Observatory and the Pawsey Supercomputing Research Centre are initiatives of the Australian Government, with support from the Government of Western Australia and the Science and Industry Endowment Fund.

The Dunlap Institute is funded through an endowment established by the David Dunlap family and the University of Toronto. B.M.G. acknowledges the support of the Natural Sciences and Engineering Research Council of Canada (NSERC) through grants RGPIN-2015-05948 and RGPIN-2022-03163, and of the Canada Research Chairs program. N.M.Mc-G. acknowledges funding from the Australian Research Council in the form of DP190101571 and FL210100039. C.S.A. acknowledges funding from the Australian Research Council in the form of FT240100498. SPO and DAL acknowledge support from the Comunidad de Madrid Atracci\'on de Talento program via grant 2022-T1/TIC-23797, and grant PID2023-146372OB-I00 funded by MICIU/AEI/10.13039/501100011033 and by ERDF, EU. TA thanks the financial support from the National SKA Program of China (2022SKA0120102). AB acknowledges financial support from the INAF initiative ``IAF Astronomy Fellowships in Italy'' (grant name MEGASKAT). SH acknowledges funding from the European Research Council (ERC) under the European Union's Horizon 2020 research and innovation programme (Grant
agreement No. 101055318). SLJ acknowledges the support of a UKRI Frontiers Research Grant (EP/X026639/1), which was selected by the European Research Council, and the STFC consolidated grants ST/S000488/1 and ST/W000903/1. RK and DL acknowledge the support of NSERC, funding reference numbers RGPIN-2020-
04853 and RGPIN-2018-03774. SL acknowledges funding from the  Australian Research Council through grant DP200100784. CJR acknowledges financial support from the German Science Foundation DFG, via the Collaborative Research Center SFB1491 `Cosmic Interacting Matters --- From Source to Signal'. A.~S.~acknowledges support from the Australian Research Council's Discovery Early Career Researcher Award (DECRA, project~DE250100003). XS is supported by the National SKA Program of China (2022SKA0120101).

The Canadian Initiative for Radio Astronomy Data Analysis (CIRADA) is funded by a grant from the Canada Foundation for Innovation 2017 Innovation Fund (Project 35999) and by the Provinces of Ontario, British Columbia, Alberta, Manitoba and Quebec, in collaboration with the National Research Council of Canada, the US National Radio Astronomy Observatory and Australia's Commonwealth Scientific and Industrial Research Organisation.  

This research used the facilities of the Canadian Astronomy Data Centre operated by the National Research Council of Canada with the support of the Canadian Space Agency, and was enabled in part by support provided by Compute Ontario (\text{https://www.computeontario.ca/}) and the Digital Research Alliance of Canada (\text{https://alliancecan.ca}).
This research has made use of NASA's Astrophysics Data System, and of adstex (\text{https://github.com/yymao/adstex})

The Digitized Sky Surveys were produced at the Space Telescope Science Institute under U.S. Government grant NAG W-2166. The images of these surveys are based on photographic data obtained using the Oschin Schmidt Telescope on Palomar Mountain and the UK Schmidt Telescope. The plates were processed into the present compressed digital form with the permission of these institutions. 

\end{acknowledgement}

\bibliography{possum_survey}

\onecolumn
\printaffiliations

\end{document}